\documentclass{cpamart1}     
\received{Month 200X}       
\volume{000}
\startingpage{1}                      

\authorheadline{R. Kleeman B. Turkington}
\titleheadline{Nonequilibrium Burgers dynamics}

\usepackage{epsfig}
\usepackage[T1]{fontenc}
\usepackage[latin9]{inputenc}
\usepackage{graphicx}
\usepackage{caption}
\usepackage{subcaption}
\usepackage{amssymb}

\usepackage{mathptmx}

\theoremstyle{definition}

\theoremstyle{remark}

\newcommand{\R}{I\!\!R}
\newcommand{\C}{l\!\!\!C}

\newcommand{\la}{\langle}
\newcommand{\ra}{\rangle}

\newcommand{\rhotilde}{\tilde{\rho}}

\renewcommand{\hbar}{\bar{h}}

\newcommand{\lambdadot}{\dot{\lambda}}
\newcommand{\lambdahat}{\hat{\lambda}}

\newcommand{\ahat}{\hat{a}}
\newcommand{\adot}{\dot{a}}

\renewcommand{\d}{\partial}
\newcommand{\half}{\frac{1}{2}}

\renewcommand{\L}{{\mathcal L}}
\renewcommand{\H}{{\mathcal H}}

\begin{document}                        


\title{A nonequilibrium statistical model of 
spectrally truncated Burgers-Hopf dynamics}

\author{Richard Kleeman}{Courant Institute of Mathematical Sciences, New York University}
\author{Bruce E. Turkington}{Department of Mathematics and Statistics, University of Massachusetts Amherst}





\begin{abstract}

Exact spectral truncations of the unforced, inviscid Burgers-Hopf equation are
Hamiltonian systems with many degrees of freedom which exhibit 
intrinsic stochasticity and coherent scaling behavior.  For this reason recent
studies have employed these systems as prototypes to test stochastic
mode reduction strategies.  In the present paper the Burgers-Hopf dynamics truncated 
to $n$ Fourier modes is treated by a new statistical model reduction technique,
and a closed system of evolution equations for the mean values of the $m$
 lowest  modes  is derived for  $m \ll n$. 
In the reduced model the $m$-mode macrostates are associated with  
trial probability densities on the phase space of the $n$-mode microstates, 
and a cost functional is introduced to quantify the lack of fit of paths of
these densities to the Liouville equation.   
The best-fit macrodynamics is obtained by minimizing the cost functional
over paths, and the equations governing the closure are then derived from Hamilton-Jacobi theory. 
The resulting reduced equations have a fractional diffusion and modified
nonlinear interactions, and the explicit form of both are determined 
up to a single closure parameter.    The accuracy and range of validity
of this nonequilibrium closure is assessed by comparison against 
direct numerical simulations of statistical ensembles, and the predicted behavior is 
found to be well represented by the reduced equations.

\end{abstract}

\maketitle   



\tableofcontents



\section{Introduction}

Throughout the applied mathematical sciences one meets complex nonlinear dynamical
systems having large finite dimension, or infinite dimension, whose typical solutions
exhibit chaotic or turbulent behavior.  Often the same systems also reveal 
some organized structures that persist within the disordered fluctuations intrinsic to 
the dynamics.    One is then faced with the challenge of describing such systems
 by appropriate model equations that synthesize the coherent 
behavior and the fluctuations, and that properly account for the interactions between them.  
A common approach is first to devise some low-dimensional governing equations
for the coherent behavior and then to add some noise, or stochastic forcing, to
capture the effect of the fluctuations.    But if the original system is already described by
an accepted set of governing equations that are deterministic,  the question 
arises:  in what sense is the stochastic model consistent with the given complex dynamics?    
To address this question of principle, it is necessary to derive the proposed stochastic model from
the underlying deterministic dynamics by a systematic reduction technique 
\cite{chorin1998optimal,givon2004extracting,mtv2,majda2006stochastic,majda2010mathematical}.   

This general problem of model reduction for complex dynamical systems, usually governed
by nonlinear ODEs or PDEs,  is conceptually identical to the generic problem of nonequilibrium
statistical mechanics, which endeavors to connect macroscopic descriptions of systems
consisting of a huge number of interacting particles or modes to the microscopic dynamics 
of their constituents \cite{balescu1975equilibrium,katz1967principles,keizer1987statistical,luzzi2002predictive,Zub74,zwanzig2001nonequilibrium}.   
The standard approach in the field of statistical mechanics is to 
invoke projection operators that map the full phase space onto the span of some 
relevant, or resolved, dynamical variables, which define a macroscopic state.   This method
is used in the well-known  Mori-Zwanzig formalism, which applies to 
an arbitrary set of resolved variables for a general Hamiltonian system 
\cite{balescu1975equilibrium,zwanzig2001nonequilibrium,chorin2000optimal,chorin2002optimal}.      Typically these
derivations make use of near-equilibrium assumptions to define tractable projection operators,
and they lead to stochastic integro-differential equations having memory kernels 
and  non-Markovian noise.  While the fluctuation-dissipation theorem determines a key
relation between the autocorrelation of the noise and the memory kernel, the evaluation of
the kernel itself is determined by ensemble averages of the certain derived observables
propagated under a complementary projection of the microdynamics.   Consequently, it is normally
imperative to make further approximations and simplifications, most notably taking a
Markovian limit, to arrive at a useful and manageable model.     As a result, systematic reductions 
of this kind are mostly confined to dynamical systems having special structures dictated by
the methodology.   

In recent work \cite{Tur12} 
one of the authors has proposed a new approach to statistical model reduction, which 
applies to any Hamiltonian dynamics and any suitable choice 
of resolved variables, and which leads directly  from the given deterministic dynamics
to closed reduced equations without recourse to intermediate stochastic approximations.    
This method of model reduction
relies on a statistical optimization principle.   First, a parametric statistical model is adopted consisting 
of trial probability densities on phase space associated with the mean values of the
selected resolved variables.  Then the reduced dynamics are
determined by minimizing a cost functional  over paths in the statistical parameter space. 
In this way a unique statistical closure is obtained once an appropriate cost functional is specified.  
The cost functional is designed to quantify the lack-of-fit of
the path of trial densities to the underlying dynamics, and information-theoretic considerations  
suggest its form as a time-integrated, weighted, squared-norm on the residual
of the trial densities with respect to the Liouville equation.  A predicted,
or estimated, 
evolution of the resolved variables therefore corresponds to a path in the statistical parameter space 
that best fits the underlying Hamiltonian dynamics in a sense quantified by the 
cost functional.       All adjustable 
parameters in the closure appear as weights in the cost functional.  These weights, which 
encode the influence of the unresolved fluctuations on the mean resolved evolution,
are determined empirically.          

The closed reduced equations that result from this dynamical optimization procedure have 
the desirable feature that they take the generic form of governing equations for nonequilibrium
thermodynamics \cite{de2011non}.    Namely, they are systems of first-order differential equations in
the mean macrostate, and they contain a reversible term, which is a generalized
Hamiltonian vector field,  and an irreversible term, which is a gradient vector field 
\cite{ottinger2005beyond}.   Moreover,
the structured derivation of these equations implies that the reversible and irreversible terms
in the reduced equations correspond to terms in the cost function that quantify the resolved and unresolved components of the Liouville residual, respectively. 

The purpose of the present paper is to illustrate and test this statistical reduction method by
applying it to the Truncated Burgers-Hopf (TBH) equation. The untruncated version of this 
partial differential equation has long been a prototype for understanding fluid shock formation 
among other phenomena \cite{lax2005hyperbolic}. 
Its $n$-mode Fourier-Galerkin truncation, without forcing or viscosity, and for sufficiently large $n$, 
has been
proposed as a good testbed for model reduction strategies, since it arises as the truncation 
 of a familiar, one-dimensional nonlinear PDE and yet shares
the statistical properties of many important complex dynamical systems
\cite{mati00,mati01,abramov2003hamiltonian,majda2006stochastic}.  In particular, the TBH dynamics is  
chaotic and numerical simulations
show it to be ergodic and mixing.  The sharp truncation at wavenumber $n$ also leads to
backscatter of energy from high to low wavenumbers, which produces coherent behavior
in the low modes.   It has a (noncanonical) Hamiltonian structure, and its
equilibrium distribution is given by a Gibbs measure.      
In addition, it has the important statistical
property that the decorrelation time of Fourier modes varies directly
with their spatial scale \cite{mati00}, a property shared by important
fluid dynamical systems such as those relevant to geophysical applications.  
The expository article \cite{mati01} describes the properties of the model both
analytically and numerically, and the paper \cite{abramov2003hamiltonian}  details the Hamiltonian
structure, the invariants and the ensemble statistical behavior.

The abovementioned relationship between decorrelation time scale and wavenumber suggests
a natural coarse-grained description in which the resolved modes are
the lowest $m$ modes, for $m \ll n$.   But, since every mode $k$ for $m+1\le k \le n$ is unresolved
in this reduction, the separation of time scales between the resolved and unresolved
modes is not wide.   Consequently, the TBH dynamics poses a stringent test of any
closure methodology, and the efficacy of a closure for the TBH system has implications
for the the design of effective reduced descriptions of many high-dimensional, complex
dynamics systems encountered in the physical sciences.

Our main result is an explicit $m$-mode reduced dynamics which has a structure similar
to $m$-mode TBH dynamics itself, but with modified nonlinear interactions and
strong, mode-dependent dissipation.    Interestingly, the dissipation derived from the
reduction procedure is not a standard diffusion, but rather a fractional diffusion.   Furthermore,
our statistical closure on $m$ resolved modes only depends on a single adjustable parameter,
which scales the magnitude of the dissipation.     By comparing the solutions of the 
reduced equations against ensemble averages of direct numerical simulations for
$n=50$ and $m=5$, we show that the reduced equations of the statistical closure reproduce the
true statistical evolution across the $m$-mode resolved spectrum, and for large as well as small
initial conditions.

\section{TBH dynamics and statistics}

The unforced, inviscid Burgers-Hopf equation is the  PDE
\begin{equation}  \label{bh}
\frac{\d u}{\d t}  + \frac{\d}{\d x} \left(  \half u^2 \right)   \, = \, 0 \,   \;\;\;\;\;\;\;\;\;\;
      (\, x \in \R^1 \, ,   t \ge 0    \, )
\end{equation}   
for a scalar unknown $u=u(x,t)$.  This equation has been extensively investigated
as a simple prototype for the nonlinear behavior typical of the multi-dimensional
equations of hydrodynamics, especially the formation of shock
discontinuities and development of turbulence.   For these considerations it is
natural to consider the Burgers-Hopf equation with a finite viscosity and with
forcing, which may be deterministic or stochastic.    By contrast, the present work
focusses exclusively on the free dynamics of spectral (or Galerkin) truncations
the inviscid equation (\ref{bh}).   These finite-dimensional dynamical system 
have been proposed as useful test systems for coarse-graining strategies and
model reduction methodologies, which are intended to be applied to much more
complex dynamic systems \cite{mati00,mati01,abramov2003hamiltonian,majda2006stochastic}.   

We study the dynamics of the projection of $u$ onto the first $n$ modes of the
Fourier series for $u$.  For  $2\pi$-periodic functions $u(x,t)$,  $x \in [0,2\pi)$,  
the truncated series  is defined by the projection operator,
\[
( P_n  u )(x,t) \, = \, \sum_{k = -n}^n z_k(t) \, e^{ikx} \, ,  \;\;\;\;\;\;\;\; 
  \;\; z_k(t) = \frac{1}{2\pi} \int_0^{2\pi} u(x,t) \,   e^{-ikx}  \, dx \, . 
\]
The mean $z_0 =   \int_0^{2\pi} u(x,t) \, dx$ is a conserved quantity under (\ref{bh}),
and so it is set to $0$ throughout.  For a real solution $u(x,t)$, 
 $z_{-k} = z^*_k$, for $k=1, \ldots , n$.   
[$z^*$ denotes complex conjugate of $z$.]    
The state space for the truncated dynamics
with $n$ modes is therefore $\, \C^n$, and a generic state is a
point $z =(z_1, \ldots, z_n) \in \C^n$.   The governing
dynamics for $u_n(x,t) = (P_n u)(x,t)$ is determined by the Galerkin truncation of (\ref{bh}), namely, 
\begin{equation}  \label{tbh-xform}
\frac{\d u_n}{\d t}  + \frac{\d}{\d x} P_n \left(  \half u_n^2 \right)   \, = \, 0 \, .
\end{equation}  
In terms of the Fourier coefficients, $z=z(t)$, the Truncated Burger-Hopf (TBH) dynamics is
accordingly the following system of $n$ quadratically nonlinear ODEs:   
\begin{equation}  \label{tbh}
\frac{d z_k}{d t}  + \frac{ik}{2} \sum_{k_1+k_2 = k}  z_{k_1} z_{k_2}   \, = \, 0 \, ,   
   \;\;\;\;\;\;\;\;\;\;   (k=1, \ldots , n).    
\end{equation} 
where $ k_1,  k_2$ run over $\{ \pm 1, \ldots , \pm n \}$.    

The dynamics (\ref{tbh})  exactly conserves the following two quantities:
\begin{equation} \label{energy}
E \, = \, \frac{1}{4\pi}  \int_0^{2\pi} u^2 \, dx  \, = \, 
\half \sum_{k=-n}^n |z_k|^2  \, , 
\end{equation}
\begin{equation}   \label{hamiltonian}
H \, = \, \frac{1}{6\pi}  \int_0^{2\pi}  u^3 \, dx \, =\, 
\frac{1}{6} \sum_{k_1+k_2+k_3 =0 } z_{k_1} z_{k_2} z_{k_3}  \, .    
\end{equation}
The quadratic invariant, $E$, defines the total energy of the truncated system (\ref{tbh}).   
Curiously, it is the cubic invariant, $H$, not the energy, $E$, that appears in the Hamiltonian
form of the  TBH dynamics (\ref{tbh});  that is, 
\[
\frac{d z_k}{dt} \, = \, \sum_{k'} J_{k k'} \frac{\d H}{\d z_k^*} \, , \;\;\;\;\;\;\;\;
\mbox{ for}  \;\;\;\;  J_{k k'} = -ik \, \delta_{k k'} \, , 
\]
where $\delta_{k k'} =1$ if $k=k'$, and $=0$ if $k \ne k'$.    These equations have the form
of a noncanonical Hamiltonian system, or Poisson system, with cosymplectic matrix $J$
and Hamiltonian $H$, having  $n$ complex degrees of freedom.        
Nonetheless,  the invariant $H$ plays a much less prominent role in conditioning the dynamics 
than does the energy $E$, as has been documented by previous researchers \cite{mati01,abramov2003hamiltonian}.

Except for $n=1,2$ the TBH dynamics is chaotic, and for $n >20$ 
numerical simulations indicate that it is ergodic and mixing.    
We therefore adopt a statistical mechanical perspective that focuses on 
the evolution of ensembles of solutions of (\ref{tbh}), rather than individual solution trajectories.    
The Hamiltonian structure of  (\ref{tbh}) implies the Liouville property, namely, 
the invariance of phase volume,
$dz = dx_1 dy_1 \ldots dx_n dy_n$,  ($z_k = x_k + i y_k$) under the 
phase flow.   Consequently, the propagation of probability by the TBH dynamics
is represented in terms of a probability density $\rho(z,t)$ with respect to $dz$
 that is transported according to the Liouville equation, 
\begin{equation}  \label{liouville}
0 \, = \, \frac{\d \rho}{\d t} + \sum_{k=1}^n 
    \dot{x}_k  \frac{\d \rho}{\d x_k} + \dot{y}_k \frac{\d \rho}{\d y_k}  \, = \, 
   \frac{\d \rho}{\d t} + \sum_{k=-n}^n 
   \dot{z}_k  \frac{\d \rho}{\d z_k}  \, .     
\end{equation}
In principle,  $\rho(z,t)$ is completely determined by (\ref{liouville}), given
an initial density $\rho(z,0)$, and hence the expectation of any observable
$F$ on $\C^n$ at time $t>0$ is determined by 
\[
\la F (t) \, | \, \rho \, \ra  = \int_{\C^n} F(z) \, \rho(z,t) \, dz \, .
\]
[Throughout the paper $\la F | \rho \ra$ denotes expectation.]  In practice, however,
the exact density $\rho(z,t)$ evolves under the nonlinear, chaotic dynamics in an extremely
complicated way, and numerical evaluation of the expectation of any
observable with respect to $\rho(z,t)$ requires  computing an ensemble of trajectories
(\ref{tbh}) starting from a sufficiently large sample of the initial density.    This intrinsic feature
of the dynamics, which is shared by many turbulent dynamical systems, is the fundamental
reason for invoking model reduction procedures, which are devised to furnish
sufficiently accurate approximations
to those expectations without expensive sampling of the detailed dynamics.    

Our reduction technique is based on using  a convenient    
family of trial probability densities on the phase space
 to approximate the exact density evolving under (\ref{liouville}).    
 This family of nonequilibrium densities is
constructed from a fixed reference equilibrium density, which we take to be
the canonical ensemble,
\begin{equation}  \label{gibbs} 
\rho_{\beta}(z)  = \frac{ \exp ( -  \beta  E(z) )}{ \int  \exp (-  \beta  E(z) ) \, dz } \, = \, 
    \prod_{k=1}^n  \frac{\beta}{\pi} e^{ - \beta |z_k|^2}   \, ,
\end{equation}
with specified inverse temperature $\beta >0$.    We choose the first $m$ components,
$A(z) = (z_1, \ldots, z_m)$, of the microstate $z \in \C^n$ to be the resolved variables, or 
resolved modes, in the model.   Experience with
computed solutions of the TBH equation, as well as a heuristic scaling argument, 
informs this choice, because it is observed that 
the decorrelation time scales for the modes scale with $1/k$  \cite{mtv2,mtv3}.      
Accordingly, the lowest $m$ modes can be considered slow variables in the $n$-mode TBH dynamics.
and the resolved vector $A$ constitutes a natural coarse-graining of the microstate $z \in \C^n$.
We call the mean values $A$ the macrostate and denote it by $a=(a_1, \ldots , a_m)$.   
This coarse-grained description choice does not, however, ensure a wide separation
of time scales between the resolved and unresolved variables, and consequently 
the TBH dynamics presents a rather challenging test for the model reduction technique.      

The general properties of the non-equilibrium probability distributions
of this system were examined numerically in \cite{klmati02} where it was noted
that they are often quasi-normal. Motivated by this empirical observation,  
our statistical model employs the quasi-equilibrium (or quasi-canonical) densities
\begin{equation}  \label{trial-densities}
\rhotilde(z;\lambda) \, = \, 
\exp \left( \sum_{k=-m}^m \lambda_k^*z_k  -   \frac{1}{2 \beta}  |\lambda_k|^2   \right)
                                             \rho_{\beta}(z)  \, ,   
\end{equation}     
This exponential family of densities (\ref{trial-densities}) form a parametric statistical model,
the parameter vector being $\lambda=(\lambda_1, \ldots , \lambda_m) \in \C^m$ \cite{casella2001statistical,kullback1997information}.   
The convention, $\lambda_{-k} = \lambda_k^*$, for  $k=1, \ldots , n$, and $\lambda_0=0$, 
applies to the parameters.
Since $\rho_{\beta}$ is a Gaussian density, each density $\rhotilde(\lambda)$
is also Gaussian;  with respect to $\rhotilde(\lambda)$ 
the means and variances of the resolved variables are
\begin{equation}     \label{qe-mean-var}
a_k = \la \, z_k \, | \, \rhotilde(\lambda) \, \ra \, = \, \frac{1}{\beta} \,  \lambda_k \, ,  \;\;\;\;\;\;\;\;
\la \, (z_k -a_k) ( z_{k'} - a_{k'})^* \, | \,  \rhotilde(\lambda)    \, \ra \, = \,  \frac{1}{\beta} \, \delta_{k k'}  \, , 
\end{equation}
for $k , k' = \pm 1, \ldots , \pm m$.    
A highly desirable feature of the statistical model (\ref{trial-densities})
is that all the modes $z_k$ are independent and the correspondence
between the mean resolved variables $a_k$ and the statistical parameters $\lambda_k$ is linear
and decoupled.       

Other possible choices of statistical model are conceivable.  For instance,  the invariance 
of both $E$ and $H$ suggests using the more general equilibrium density
\[
\rho_{\beta, \alpha, \eta} (z) \, = \, Q(\beta, \alpha, \eta)^{-1} 
            \exp ( \, -\beta E(z)  - \eta E(z)^2  - \alpha H(z) \, ) 
\]    
with $\eta >0$;  $Q(\beta, \alpha, \eta)$ is the normalizing factor. 
This density is
a so-called Gaussian ensemble with respect to $E$, as it includes the quadratic term $E^2$ in
the exponential, which  is necessary to make the density normalizable given the presence
of the cubic invariant $H$.   But, using $\rho_{\beta, \alpha, \eta} $ as the reference
density for the statistical model $\rhotilde(\lambda)$ leads
to non-Gaussian trial densities, and hence an intractable subsequent analysis.     Fortunately, 
simulations of ensembles of solutions of the TBH dynamics have shown that the distribution
of the modes $z_k$ is very nearly Gaussian, and that $H$ has only 
a slight effect on the statistical behavior, provided that the mean value $\la H \ra$
is close enough to the relaxed value $\la H \, | \, \rho_{\beta} \ra =0\; $  \cite{mtv2,mtv3,abramov2003hamiltonian}.      
Accordingly, our statistical model and closure procedure are built from the reference density (\ref{gibbs})
that does not include $H$.   In this formulation we expect to approximate a statistical evolution 
in which the mean value of $H$ is near its relaxed value, but not an evolution from
an initial statistical state with a large mean value of $H$.      

Neither $\la E | \rhotilde(\lambda)  \ra $ nor $\la H | \rhotilde(\lambda)  \ra $ is exactly conserved 
along an arbitrary admissible path $\lambda=\lambda(t)$ in the statistical parameter space.    
It is possible to alter the formulation of the statistical model to ensure that both of these 
invariants are respected, but with the inevitable consequence that the trial densities are
non-Gaussian \cite{Tur12}.    Our main justification for allowing some variation in these dynamical
invariants is, therefore,  expediency, combined with the fact that for $n \gg m$ the relative 
variations in the means of $E$ and $H$ are small.      
 
\section{Formulating the optimization principle}   

Our $m$-mode closure of the $n$-mode TBH dynamics is derived from an optimization principle over paths,
$\lambda(t)$, in the parameter space, $\C^m$, of the statistical model  (\ref{trial-densities}).
A predicted, or estimated, evolution of the macrostate $a=a(t)$, with
$a_k (t) = \la \, z_k \, | \,  \rhotilde(\lambda(t) ) \, \ra $ for $k = 1, \ldots , m$, corresponds to
a path $\lambda(t)$ that is optimally compatible with the underlying TBH dynamics, in
the sense that $\lambda(t)$ minimizes a certain lack-of-fit cost functional over all
admissible paths.   This lack-of-fit is a metric on the Liouville residual, 
a fundamental statistic in our approach that we define to be 
\begin{equation}   \label{residual}
R \, = \,  \left( \,  \frac{\d }{\d t} + \sum_{k = -n}^n \frac{d z_k}{dt} \frac{\d }{\d z_k} \, \right) 
                    \log \rhotilde(\cdot \, ; \lambda(t))   
    \, = \, \sum_{k = -m}^m  \lambdadot_k^* ( z_k -a_k)  
                    +     \lambda_k^* \frac{d z_k}{dt} \, .
\end{equation}   
Here $d z_k/dt$ is given by the governing TBH dynamics (\ref{tbh}),
and $\lambdadot_k = d \lambda_k /dt$.   If $\rhotilde( \cdot \, ; \lambda(t))$ is replaced
by an exact solution $\rho(\cdot \, , t)$ of (\ref{liouville}) in (\ref{residual}), then $R=0$
identically on the phase space $\C^n$.     In \cite{Tur12} it is shown that the
statistic $R=R(z;\lambda,\lambdadot)$ may be interpreted as the local rate of information loss 
at the sample point $z$ due to reduction via the statistical model (\ref{trial-densities}).   
The significance of $R$ is also revealed by the family of identities
\[
\frac{d}{dt}  \la \, F \, | \, \rhotilde(\lambda(t)) \, \ra \, - \,  \la \, \dot{F} \, | \, \rhotilde(\lambda(t)) \, \ra \,
 = \,  \la \, FR \, | \, \rhotilde(\lambda(t)) \, \ra \, , 
\]  
which hold for every observable $F$ on $\C^n$, with $\d F / \d t=0$;   
here, $\dot{F} = \{ F, H \}$ denotes the 
Poisson bracket of $F$ with $H$, so that $d F(z(t))/dt = \dot{F} (z(t))$ on solutions $z=z(t)$
of (\ref{tbh}).   Since the mean of $R$ with respect to $\rhotilde(\lambda)$ is zero, 
$\la R | \rhotilde \ra = 0$, the covariance between an observable $F$ and the Liouville
residual $R$ quantifies the deficiency of the path of trial densities $\rhotilde(\lambda(t))$
to propagate the expectation of $F$.    This family of identities over test observables
 furnishes a natural linear
structure for defining the lack-of-fit of any admissible path $\lambda(t)$ to the underlying
dynamics, and motivates the use of a weighted $L^2(\C^n, \rhotilde(\lambda))$ squared norm as the 
lack-of-fit cost function.   

The weights in the lack-of-fit cost function relate to the resolved and unresolved
subspaces of $L^2(\C^n, \rhotilde(\lambda))$.  The mean-centered resolved variables
\begin{equation}  \label{score}
U_k  = z_k - a_k \, = \, \frac{ \d }{\d \lambda_k } \log \rhotilde ( \cdot \, ; \lambda )  \,  \;\;\;\;\;\;\;\;\;\;
( \, k = 1, \ldots , m \, ) \, , 
\end{equation}
span the resolved subspace,  and,    
by (\ref{qe-mean-var}), they are uncorrelated Gaussian variables, 
$\la U_k U_{k'}^* \ra =  \beta^{-1} \delta_{k k'}$, for all $\lambda \in \C^n$.   The
orthogonal projection of $L^2(\C^n, \rhotilde)$ onto the resolved subspace is
\begin{equation}  \label{proj-P-def}
P_A F \, = \, \beta \sum_{k = -m}^m \la \, F U_k^* \, | \, \rhotilde \,  \ra \, U_k    \, ,    
\end{equation}
and the complementary projection onto the unresolved subspace (the orthogonal
complement of the resolved subspace) is $Q_A = I - P_A$.   
The projections of the Liouville residual $R=R(\lambda, \lambdadot)$ onto its resolved and unresolved components are calculated to be  
\begin{equation}  \label{PR}
P_A R \, = \,   \beta \sum_{k = -m}^m 
       \left[ \, \adot_k +  \frac{ik}{2} \sum_{k_1 +k_2 = k} a_{k_1} a_{k_2}  \, \right]^* U_k 
\end{equation}
\begin{equation}   \label{QR}   
Q_A R  \,=\,  \frac{ -i \beta}{2} \sum_{k = -m}^m k a_k^* \sum_{k_1 +k_2 = k} U_{k_1} U_{k_2}        
     \,=\,  \frac{ i \beta}{2}   \sum_{k_1 +k_2 + k_3 =0} k_3 a_{k_3} U_{k_1} U_{k_2}  \, . 
\end{equation}
Here, $\adot_k = d a_k / dt = \lambdadot_k / \beta$.    
Since $a_k=0$ for $|k| > m$,  the wavenumber indices $k_1, k_2$ in (\ref{PR})
run over $\{ \pm 1, \ldots , \pm m  \}$, unlike the nonlinear term in the TBH
dynamics, in which all modes in $\{ \pm 1, \ldots , \pm n  \}$ interact 
with the resolved modes.   By contrast, $k_1, k_2$ in (\ref{QR}) run over $\{ \pm 1, \ldots , \pm n  \}$,
and the definition of $U_k$  is extended for  $|k| > m$ to be simply $U_k = z_k$.   

The identities  (\ref{PR}) and (\ref{QR}) make use of the following
identities for the Gaussian scores variables:
\begin{eqnarray}   \label{3and4moments}   
\la \, U_{k_1} U_{k_2} U_{k_3} \, | \, \rhotilde \, \ra &=& 0 \, , \\   \nonumber
\la \, U_{k_1} U_{k_2} U_{k'_1}^* U_{k'_2}^*\, | \, \rhotilde \, \ra & = &  \frac{1}{\beta^2} 
        [ \,  \delta_{k_1 k'_1} \delta_{k_2 k'_2} + \delta_{k_1 k'_2}\delta_{k_2 k'_1} \, ]   \, .  \nonumber
\end{eqnarray}
The  calculations (\ref{PR}) and (\ref{QR}) are summarized as follows.   
In light of (\ref{residual}), 
\[
P_A R \, = \,  \sum_{k = -m}^m  \lambdadot_k^* U_k  
   +  \lambda_k^* \beta  \sum_{k' = -m}^m \la \, \frac{d z_k}{dt} \, U_{k'}^* \,  \ra \, U_{k'} \, .
\]
Substituting (\ref{tbh}) and using 
$ \beta \la \, z_{k_1}z_{k_2} U_{k'}^* \, \ra = 
         a_{k_1}  \delta_{k_2 k' }  +  a_{k_2} \delta_{k_1 k'}   \,   $, 
which is implied by (\ref{3and4moments}),   results in 
\begin{eqnarray*}
P_A R & = &  \sum_{k = -m}^m  \lambdadot_k^* U_k  
   +  \lambda_k^* \, ik \sum_{k' = -m}^m a_{k-k'}  U_{k'}    \\
     & =&  \sum_{k = -m}^m  \left\{ \lambdadot_k^*  
              - i  \sum_{k' = -m}^m k' \lambda_{k'}^*  a_{k-k'}^*   \right\} U_{k}    \\  
     & =&  \sum_{k = -m}^m  \left\{ \lambdadot_k^*  
             +  i  \sum_{k_1+k_2=k}  k_1 \lambda_{k_1}  a_{k_2}   \right\}^* U_{k}  \, .   \\      
\end{eqnarray*}
Putting $\lambda_k=\beta a_k$ and using the symmetry between $k_1$ and $k_2$ 
then produces the desired identity (\ref{PR}).   
In a similar manner,
\begin{eqnarray*}
Q_A R & = &  R \, - \, P_A R  \\  
             & =&  \sum_{k = -m}^m 
     \lambda_k^* \left[ \frac{d z_k}{dt} +  \, ik \sum_{k' = -m}^m a_{k-k'} \right] U_{k'}    \\ 
     & =& - \frac{i}{2}  \sum_{k = -m}^m   
         k  \lambda_k^* \sum_{k_1+k_2=k}  z_{k_1}  z_{k_2}   - a_{k_1} U_{k_2} - a_{k_2} U_{k_1}   \\  
     & =&   - \frac{i}{2}  \sum_{k = -m}^m   
         k  \lambda_k^* \sum_{k_1+k_2=k}  U_{k_1}  U_{k_2}   - a_{k_1} a_{k_2}    \, ,      
\end{eqnarray*}
using the symmetry between $k_1$ and $k_2$ in the third equality.
 The constant term in the last   expression vanishes, because 
\[
 \sum_{k = -m}^m  k  \lambda_k^* \sum_{k_1+k_2=k}  a_{k_1} a_{k_2}   \, = \, 
 - \beta \sum_{k_1+k_2+k_3 = 0}   k_3 a_{k_1} a_{k_2} a_{k_3}  \, = \, 0 \, ,   
\]
and hence the desired identities (\ref{QR}) follow.

The cost function for our optimization principle is declared to be
\begin{equation}  \label{cost-fn-def}
\L(\lambda, \lambdadot)  \, = \, \half  \la \, [ P_A R (\lambda,\lambdadot) ]^2  \, | \, \rhotilde(\lambda) \, \ra
     +    \half \la \, [ W Q_A R (\lambda,\lambdadot) ]^2  \, | \, \rhotilde(\lambda) \, \ra  \, ,
\end{equation}
in which $W$ is a linear operator on $L^2(\C^n, \rhotilde)$ satisfying $W P_A = P_A W$.
We refer to $W$ as the weight operator, because it weights the contributions of the
unresolved component, $Q_A R$, of the  Liouville residual $R$ relative to the
(unit weighted) resolved component, $P_A R$.     
The requirement that $W$ commutes with the projection $P_A$
implies that $W$ also commutes with $Q_A$ and hence that $W$ takes the unresolved subspace into itself.  
The inclusion of $W$ in the cost function (\ref{cost-fn-def})  
gives our best-fit closure the character of a weighted least-squares approximation over paths $\lambda(t)$. 

With this cost function in hand, we are now able to formulate the optimization principle
that defines our statistical closure.   In this section we present the stationary
version of the principle, which is simpler to describe and motivate.   A nonstationary version,
which is needed when comparing the predictions of the closure with direct numerical
simulations, is given in Section 6.    

The best-fit closure scheme is based on the dynamical minimization problem
\begin{equation}    \label{value-fn}
v(\lambda^0) \, = \,   \min_{\lambda(t_0)=\lambda^0} \; \int_{t_0}^{+ \infty}  
        \L(\lambda, \lambdadot)  \, dt \,  , 
 \end{equation}   
in which the admissible paths 
$\lambda(t), \; t_0 \le t < + \infty \, $, in the configuration space of the
statistical model are constrained to start  at $\lambda^0 \in \C^n$ at time $t_0$.    
 In optimization and control theory, $v(\lambda^0)$ is called
  the value function for the minimization problem (\ref{value-fn}) \cite{bryson1975applied,evans1998partial,fleming1975deterministic}.   
  Since $\L(\lambda, \lambdadot)$ is independent of $t$, and the integration
extends to infinity in time, $v(\lambda)$ is time-independent,
and the initial time $t_0$ may be shifted to $0$ in (\ref{value-fn}).   
  
By analogy to analytical mechanics, one may regard (\ref{value-fn}) as a principle of
least action for the ``Lagrangian'' $\L(\lambda, \lambdadot)$ and interpret the
first member  in (\ref{cost-fn-compact}) as 
its ``kinetic'' term and the second member as its ``potential'' term \cite{arnol1989mathematical,gelfand2000calculus,lanczos1986variational}.
The kinetic
term is a positive-definite quadratic form in the generalized velocities $\lambdadot$,
and it is entirely determined by the structure of the resolved variables and trial densities.   
By contrast, the potential term  embodies the influence of the unresolved
variables on the resolved variables and involves the weight operator $W$, which 
contains all the adjustable closure parameters.   

Each extremal path $\lambdahat(t)$, $t_0 \le t < +\infty$,  for (\ref{value-fn}) corresponds to 
an  evolving trial density $\rhotilde(\, \cdot \, ; \lambdahat(t))$ that is best-fit 
to the Liouville equation in the sense that 
the time-integral of the lack-of-fit cost function is minimized.   Finiteness of the value
function implies that $\lambda(t) \rightarrow 0$ as $t \rightarrow +\infty$, meaning that
the best-fit path connects the given initial state $\lambda^0$ to equilibrium $\lambda^{eq} =0$.  
Hence, these extremal paths
model the relaxation of the mean resolved vector $\ahat(t) = \la A \, | \rhotilde(\lambdahat(t)) \ra$,
from a given initial state $a^0 = \la A \, | \rhotilde(\lambda^0) \ra$.      

It is important to emphasize that the weight operator $W$ is an independent ingredient
in our closure strategy and that $W$ includes all the empirical parameters in the 
best-fit closure.    It is unavoidable that some empirical ingredient should enter into the
definition of such a closure, because the defining optimization principle  does not refer to the 
autocorrelations of the unresolved modes under the full dynamics, or the projected dynamics
orthogonal to the resolved subspace.    As is well-known in nonequilibrium statistical mechanics,
the transport properties of the complex system, as well as the memory kernels in its projected dynamics, are 
expressible in terms of such autocorrelations \cite{balescu1975equilibrium,Cha87,Zub74,zwanzig2001nonequilibrium}.   
In this light our closure strategy of adopting a time-integrated, weighted, least-squares 
approximation
is a practical expedient to avoid the expensive computation of such autocorrelations, which 
require the propagation of ensembles of solutions of the fully resolved dynamics.  
Our approach instead quantifies the influence of the unresolved modes on the resolved modes by
an appropriately chosen weight operator $W$ which depends on some adjustable parameters
that must be tuned empirically.    

In this general formulation of the optimization principle, the weight operator $W$ is not restricted to any particular form.  Moreover, the closed reduced equations satisfied by its extremals have
a generic thermodynamic format and associated properties for any admissible choice of $W$ \cite{Tur12}.   
In any concrete
implementation of the best-fit approach, however, it is essential to discern a convenient form for $W$ from
the structure of the underlying dynamics, the resolved variables and the unresolved Liouville residual.  
It is also desirable to choose a weight operator that contains as few adjustable
parameters as possible, given the required level of fidelity of the closure approximation to the true statistical
dynamics.    In the Burgers-Hopf dynamics the quadratic nonlinearity, typical of hydrodynamics, affords a particularly
simple and efficacious choice of $W$, as is described in the next section.

\section{Deriving the closed reduced dynamics}

First, let us consider the implications of the trivial choice for the weight operator in (\ref{cost-fn-def}), 
namely, $W=0$.    In that situation
the solution of the minimization problem (\ref{value-fn}) is complete determined by 
 setting $P_A R = 0$ at each instant of time.  The resolved vector,   
 $a= (a_1(t), \ldots , a_m(t)) = ( \la z_1|\rhotilde \ra , \ldots , \la z_m|\rhotilde \ra ) $,  then satisfies 
\[
\frac{d a_k}{dt} +  \frac{ik}{2} \sum_{k_1 +k_2 = k} a_{k_1} a_{k_2}  \, = 0   
\;\;\;\;\;\;\;\;\;\;\;\;\;\;  (k = 1, \ldots , m) \, , 
\]
meaning that the reduced dynamics coincides with the spectral truncation to $m$ modes of the
Burgers-Hopf equation.   Not only are the $m$-mode truncations of $E$ and $H$ invariant
under this dynamics, but  the entropy,  
\begin{equation}   \label{entropy}  
s \, = \, - \la \, \log \rhotilde \, | \, \rhotilde \, \ra  \, = \, - \frac{\beta}{2} \sum_{ k = -m}^m |a_k|^2   \, , 
\end{equation}   
is also invariant; indeed, the entropy production is 
\[
\frac{ds}{dt}  \,  = \,  - \, \beta \sum_{ k = -m}^m a_k^* \, \frac{ d a_k}{dt}  
                   \; = \;  \frac{i \beta}{2}   \sum_{k_1 +k_2 + k_3 = 0 } k_3 a_{k_1}a_{k_2}a_{k_3 }
                        \, = 0 \, .
\]
This naive closure is therefore adiabatic, in that it suppresses interactions between
the resolved  modes and the unresolved modes, those interactions being 
the source of entropy production and dissipation in a proper reduced dynamics.   

Now, we seek an effective choice of $W$ that appropriately quantifies the cost of 
the unresolved component of the Liouville residual as well as the resolved component. 
An examination of (\ref{QR}) reveals that unresolved residual, $Q_A R$, 
 is a linear combination of the products $U_{k_1} U_{k_2}$
over $k_1, k_2 = \pm 1, \ldots , \pm n$.   This structure is a consequence of the quadratic nonlinearity
of the TBH dynamics.     Since these products form an orthogonal basis for the unresolved subspace,
it is natural and convenient to construct the desired weight operator in terms of them.   
Accordingly, we take $W$ to be diagonal with respect to this basis and we set 
\begin{equation}  \label{weight-op}
W ( U_{k_1} U_{k_2} ) \, = \, \epsilon_{k_1,k_2} \, ( U_{k_1} U_{k_2} )  \, , 
\end{equation}
for some sequence of constants $ \epsilon_{k_1,k_2} \ge 0$, symmetric with respect to interchanging
$k_1$ with $k_2$.     
The ``potential'' term in the cost function (\ref{cost-fn-def}) is then  
\begin{eqnarray}   \label{closure-potential}
    \half \la \, [ W Q_A R (\lambda,\lambdadot) ]^2  \, | \, \rhotilde(\lambda) \, \ra 
                   &=&  \frac{1}{4}  \sum_{k=-m}^m  k^2 | a_k|^2  \cdot
                             \sum_{k_1 + k_2 = k} \epsilon_{k_1, k_2}^2  
                                    \la \,  U_{k_1} U_{k_2} U_{k_1}^* U_{k_2}^*   \, \ra \nonumber \\  
                    &=&  \frac{1}{2\beta^2} \sum_{k=-m}^m    \gamma_k \, k^2 | a_k|^2  \, ,      
\end{eqnarray}
with 
\begin{equation}  \label{gamma-k}   
\gamma_k \, = \,   \half  \sum_{k_1 + k_2 = k} \epsilon_{k_1, k_2}^2    \;\;\;\;\;\;\;\;\;\;\;\;
  ( \, k = 1, \ldots , m \, )  \, .    
\end{equation}
The sums in (\ref{closure-potential}) and (\ref{gamma-k}) extend over $1 \le |k_1|, |k_2| \le n$.   
The calculation in (\ref{closure-potential}) uses the fourth moments in (\ref{3and4moments}).  
We thus arrive at the explicit expression for the cost function,
\begin{equation}   \label{cost-fn}
\L(\lambda, \lambdadot) \, = \, \half  \sum_{k=-m}^m
        \frac{1}{\beta} \left| \,   \lambdadot_k + 
                  \frac{ik}{2 \beta} \sum_{k_1 +k_2 = k} \lambda_{k_1} \lambda_{k_2}  \, \right|^2
               \, + \,   \frac{ \gamma_k \, k^2}{\beta^2}  \, | \lambda_k |^2  \, ,   
\end{equation}
which follows from (\ref{cost-fn-def}) by (\ref{PR}) and (\ref{closure-potential}),
together with the simple relations given in (\ref{qe-mean-var}).    

We notice that  the weights $\epsilon_{k_1,k_2}$ defining $W$ enter 
the cost function $\L$ only though the combinations $\gamma_1, \ldots, \gamma_m$ in (\ref{gamma-k}).  
While it is possible to retain these $m$ adjustable parameters in the closure, their form
as sums over the weights  $\epsilon_{k_1,k_2}$ suggests a further simplification.  Namely, 
for $ 1 \le k \le m \ll n$, the $\gamma_k$ in (\ref{gamma-k}) are approximately independent
of $k$, whenever the weight sequence $ \epsilon_{k_1,k_2}$ is chosen to be
bounded and slowly varying over the range $1 \le |k_1|, |k_2| \le n$.   On this basis it is reasonable
to set $\gamma_k = \gamma >0$, for $k=1, \ldots, m$, and to adopt a cost function
 with a single adjustable parameter, $\gamma$.     
We confine ourselves to this simplified formulation throughout the present paper.   
This choice of cost function is also justified {\it a posteriori} by our subsequent analysis, which  shows that the 
scaling of the modal dissipation rates with $|k|$ implied by
a constant $\gamma$  agrees  with the direct numerical simulations presented in Sections 7 and 8  
as well as the  formal dimensional argument summarized in Section 5.       
Indeed, we find that the TBH dynamics is an example of a complex system in which
the effect of unresolved modal fluctuations on the resolved modes is efficiently
and accurately modeled by a lack-of-fit cost function having a single, empirically-tuned, closure parameter.

The equations satisfied by the extremal paths for (\ref{value-fn}) are found by the techniques of 
the calculus of variations \cite{bryson1975applied,fleming1975deterministic,evans1998partial,gelfand2000calculus,sagan1992introduction}.    
  In particular, Hamilton-Jacobi theory furnishes the
 the closed reduced equations for the statistical model in terms of the value
 function $v(\lambda)$.    To this end 
 we write the cost-function (\ref{cost-fn}) in the more compact form,
\begin{equation}   \label{cost-fn-compact}
\L(\lambda, \lambdadot) \, = \, \sum_{k=1}^m
        \frac{1}{\beta} \left| \,   \lambdadot_k  - \beta f_k(\lambda) \, \right|^2 \, + \,  
                                 \frac{ \gamma \, k^2}{\beta^2}  \, | \lambda_k |^2  \, , 
\end{equation}
introducing the shorthand notation
\begin{equation}    \label{f}
f_k(\lambda) \, = \, 
                -  \frac{ik}{2 \beta^2} \sum_{k_1 +k_2 = k} \lambda_{k_1} \lambda_{k_2}  \, . 
\end{equation}
and recalling  that $\lambda_{-k} = \lambda_k^*$.    We form the Hamiltonian $\H(\lambda,\mu)$
conjugate to the Lagrangian $\L(\lambda, \lambdadot)$ by  taking the Legendre transform of  $\L$:
 \begin{eqnarray}  \label{legendre}
\mu_k &= &   \frac{ \d \L }{\d \lambdadot_k^*} \, = \, \frac{1}{\beta}   \lambdadot_k - f_k(\lambda)  \, ,  \\
  \nonumber 
\H(\lambda, \mu) &=&  \sum_{k=1}^m \mu_k ^*\lambdadot_k  +   \lambdadot_k^*\mu_k  
     - \L(\lambda,\lambdadot)   \\  \nonumber 
        &=&  \sum_{k=1}^m  \beta |\mu|^2 + \beta [ \, \mu_k^*f_k(\lambda)    + f_k(\lambda)^* \mu_k  \, ]
                 - \frac{\gamma}{\beta^2} k^2 | \lambda_k |^2     \, .   
\end{eqnarray}
[That these complex expressions are appropriate complexifications of 
the real Legendre transform is shown in the Appendix.]  

The value function, $v(\lambda)$, in  (\ref{value-fn}), which
is analogous to an action integral or Hamilton principal function,  satisfies the
stationary Hamilton-Jacobi equation:
\begin{equation}  \label{HJ}
\H \left( \lambda, \, - \frac{\d v}{\d \lambda^*} \right) \, = \, 0 \, ,
\end{equation}
According to Hamilton-Jacobi theory, 
the conjugate variable $\mu = \hat{\mu}(t)$ along an extremal
path $\lambda = \lambdahat(t)$ is given by the relation 
\begin{equation}   \label{conjugate-relation}
\hat{\mu} \, = \,   - \frac{\d v}{\d \lambda^*}(\lambdahat)      \, . 
\end{equation}  
This key relation closes the reduced dynamics along the extremal path, since 
together with (\ref{legendre}) it yields the system of first-order ODEs 
\begin{equation}  \label{closed-reduced-1}  
 \frac{d \ahat_k}{dt}  \, = \, f_k(\lambdahat)  \, - \, \frac{\d v}{\d \lambda_k^*}(\lambdahat)  \, ,
 \;\;\;\;\;\;\;\;\;\;\;\;  ( \,   k = 1, \ldots , m   \, ) 
\end{equation}   
recalling that $\ahat_k = \lambdahat_k / \beta$.   When expressed entirely in 
terms of the mean resolved variables, the closed reduced equations are
\begin{equation}   \label{closed-reduced-2}  
 \frac{d \ahat_k}{dt} + \frac{ik}{2} \sum_{k_1 +k_2 = k} \ahat_{k_1} \ahat_{k_2} 
    \, = \,   -  \frac{\d v}{\d \lambda_k^*}(\beta \ahat)    \, .  
\end{equation}    
   
The left-hand side of this system of equations is recognizable as the $m$-mode
TBH dynamics.   The adiabatic closure mentioned earlier in this section
is obtained by setting $v=0$ identically in (\ref{closed-reduced-2}).      
The presence of the gradient vector field of $v$ on the right hand side provides the dissipation,
or irreversibility, of the closed reduced dynamics.   Indeed,
the relation (\ref{conjugate-relation}) has the interpretation that the conjugate vector $\mu$  
to the statistical state vector $\lambda$ represents the irreversible part of the flux of
the mean resolved vector $a$.    

The best-fit closure (\ref{closed-reduced-1},\ref{closed-reduced-2}) is completed once  the value function
(\ref{value-fn}) is determined explicitly, at least to some suitable approximation.    This calculation is the content of the next section.    

\section{Approximating the value function}    

Before presenting the nonlinear analysis needed to approximate the value
function, we exhibit the first-order, linear approximation in the near-equilibrium regime, 
which can be obtained by elementary methods.     We consider
the relaxation of a small initial disturbance from equilibrium, and so we retain
 only the quadratic terms in the cost function (\ref{cost-fn-compact}), 
which becomes 
 \[
 \L(\lambda, \lambdadot)  \, = \, \sum_{k=1}^m
        \frac{1}{\beta} | \,   \lambdadot_k |^2 \, + \,  
                                 \frac{ \gamma \, k^2}{\beta^2}  \, | \lambda_k |^2  \, .
\]
Each extremal, $\lambdahat(t)$, solves the Euler-Lagrange equations for (\ref{value-fn})
for this integrand,  
\[
 - \frac{ d^2 \lambdahat_k}{ dt^2}   + \frac{\gamma k ^2}{\beta}  \lambdahat_k \, = \, 0 \, .   
\]
Since the admissible paths connect the initial state  to equilibrium as
$t \rightarrow +\infty$, the relevant extremals are the exponentially decaying solutions,
\[
\lambdahat_k(t) =  \lambda_k(0) \exp \left(  -  \sqrt{\frac{\gamma}{\beta}} \, |k| t  \right).  
\]
The near-equilibrium closed reduced equations satisfied by these extremals are
\begin{equation}  \label{simple-cr}
\frac{ d \ahat_k}{ dt}  \, = \, -  \sqrt{\frac{\gamma}{\beta}} \, |k| \,   \ahat_k    \, ,
\end{equation} 
and the  value function evaluated on these extremals is  
\[
v(\lambda)  \, = \,    \sqrt{\frac{\gamma}{\beta^3}} \, \sum_{k=1}^m k \, |\lambda_k|^2 \, .
\]  

The elementary result (\ref{simple-cr}) is consistent with (\ref{closed-reduced-1}), 
since $f=0$ in this approximation. 
Moreover, the same result is anticipated by a heuristic scaling argument in \cite{mati00,mati01}.   
Phenomenological reasoning about the transfer of energy in the turbulent TBH system
suggests forming  an eddy turnover time, $T_k$, for the $k$-th mode  from
the wavenumber, $k$, and the energy per mode in statistical equilibrium, $1/\beta$.  
The only dimensionally consistent time scale is $T_k \sim \sqrt{\beta}/k$.   
It is well confirmed by direct numerical simulations of the TBH dynamics in \cite{mati00,mati01} that the
equilibrium autocorrelations of the modes, $z_k$, exhibit the $k$-dependent decay rate
proportional to $T_k^{-1}$ across the spectrum, thus validating this scaling argument.
To relate our basic prediction (\ref{simple-cr}) to this scaling theory and its
validating numerical simulations we rely on linear
response theory, which holds for relaxation near equilibrium and connects the nonequilibrium mean
value of the resolved variables $A_k$ to their equilibrium autocorrelations \cite{Cha87,katz1967principles,zwanzig2001nonequilibrium}.
Precisely, the ensemble mean of $A_k(t)$, the  resolved variable propagated under the phase flow
for time $t$,
with respect to an initial quasi-equilibrium density  $\rhotilde(\lambda^0)$  for small $|\lambda^0|$ is 
calculated up to linear-response approximations to be
\[
\la A_k(t) \, | \, \rhotilde(\lambda^0) \ra \, \approx 
\la A_k(t) \, [  1 + \sum_{k'=-m}^m ( \lambda_{k'}^0)^* A_{k'} ] \ra_{eq}
 =  \sum_{k'=-m}^m  \la A_k(t) A_{k'}^* \ra_{eq}  \, \lambda_{k'}^0   \, .   
\]
Thus the time scale for decay of $\la A_k(t) \, | \, \rhotilde(\lambda^0) \ra$ is dictated by the time scalings
of the decaying autocorrelations $  \la A_k(t) A_{k'}^* \ra_{eq}  $.   In light of the robust scaling behavior
of the autocorrelations, the decay rates in (\ref{simple-cr}) are necessarily proportional to $|k|/\sqrt{\beta}$.
This correspondence also validates the choice of the weight operator in the cost function and the
further simplification to a single closure parameter $\gamma$, independent
of $k$.     

We now turn to the nonlinear problem of determining an approximation to $v(\lambda)$
that is valid for larger amplitudes $|\lambda|$.   We seek an solution that is accurate
up to cubic terms in $\lambda$, so that the resulting gradient term in (\ref{closed-reduced-2})
contributes quadratically-accurate nonlinear terms.    Accordingly, we submit a
third-order Taylor expansion for $v$ to the Hamilton-Jacobi equation (\ref{HJ}):
\begin{equation}   \label{taylor} 
v(\lambda) \, = \, \sum_{k=1}^m M_k |\lambda_k|^2   \, + \, 
      \sum_{k_1,k_2,k_3}  N_{k_1k_2k_3} \lambda_{k_1}\lambda_{k_2}\lambda_{k_2}
    \, + \ldots  \, . 
\end{equation}  
The constant and first-order terms in (\ref{taylor}) vanish, since $v(0)=0 = \d v/\d \lambda(0)$. 
The quadratic terms  in the expansion (\ref{taylor}) already have the special form
anticipated by the foregoing linear analysis,  with real and positive coefficients 
$M_k$.   The indices $k_1, k_2, k_3$ for the cubic terms run over $\{ \pm 1, \ldots , \pm m\}$,  
and the  complex coefficients
$N_{k_1k_2k_3}$  are assumed to be symmetric under permutation of $k_1, k_2, k_3$
and to satisfy $ N_{-k_1,-k_2,-k_3} = N_{k_1k_2k_3}^*$, since   $v(\lambda)$ is a real-valued function.  These requirements supplemented the convention that 
 $\lambda_{-k} = \lambda_k^*$ and  $\mu_{-k}=\mu_k^*$.   The reader is referred to the Appendix 
for a summary of the appropriate  form of Taylor's theorem for functions of several complex variables.

The Hamiltonian in  (\ref{HJ}), when written in full using (\ref{f}), is 
\begin{equation}   \label{ham-full}  
\H(\lambda,\mu) \, = \,  \sum_{k=1}^m  \beta |\mu_k|^2  - \frac{\gamma}{\beta^2} k^2 | \lambda_k |^2   
 \; + \, \frac{i}{2\beta} \sum_{k_1+k_2+k_3=0}  k_3  \lambda_{k_1}\lambda_{k_2}\mu_{k_3}  \, . 
\end{equation}   
The leading term in (\ref{HJ}) is evaluated as follows:
\begin{eqnarray}   \label{leading}
\sum_{k=1}^m \left| \frac{\d v}{\d \lambda_k^*}  \right|^2   &=&   
\sum_{k=1}^m \frac{\d v}{\d \lambda_k}  \frac{\d v}{\d \lambda_{-k}}   \\ \nonumber
&=&   \sum_{k=1}^m M_k^2 |\lambda_k|^2  \, + \,  
              \sum_{k_1,k_2,k_3}  N_{k_1k_2k_3}  \sum_{k=1}^m M_k 
                   \left( \lambda_k \frac{\d}{\d \lambda_k}  +  \lambda_{-k} \frac{\d}{\d \lambda_{-k} } \right) 
                        \lambda_{k_1}\lambda_{k_2}\lambda_{k_3}   \\ \nonumber
 &=&      \sum_{k=1}^m M_k^2 |\lambda_k|^2  \, + \,      
                  \sum_{k_1,k_2,k_3}  N_{k_1k_2k_3}  (  M_{k_1} +   M_{k_2} + M_{k_3}  )        
                     \lambda_{k_1}\lambda_{k_2}\lambda_{k_3}     \; .   
\end{eqnarray}
Substituting $ \d v / \d \lambda_k^* = M_k \lambda_k + O(|\lambda|^2)$ into (\ref{HJ}), 
and using (\ref{leading}), yields  the Hamilton-Jacobi equation up to third order:  
\begin{eqnarray}   \label{HJ-cubic} 
 & & \sum_{k=1}^m  \beta M_k^2 |\lambda_k|^2  - \frac{\gamma}{\beta^2} k^2 | \lambda_k |^2   \, + \, 
       \beta  \sum_{k_1,k_2,k_3}  N_{k_1k_2k_3}  (  M_{k_1} +   M_{k_2} + M_{k_3}  )        
                     \lambda_{k_1}\lambda_{k_2}\lambda_{k_3}   \\   \nonumber
 & &   \;\;\;\;\;\;\;\;\;\;    
  - \, \frac{i}{2\beta} \sum_{k_1+k_2+k_3=0}  k_3 M_{k_3} \lambda_{k_1}\lambda_{k_2}\lambda_{k_2} 
   \;  =  \, 0  \, . 
\end{eqnarray}
Equating the quadratic terms in (\ref{HJ-cubic}) produces the anticipated coefficients,
\begin{equation}  \label{Mk} 
M_k = \sqrt{ \frac{\gamma}{\beta^3}  }  |k|     \;\;\;\;\;\;\;\;\;\;   (k = 1, \ldots , m) \, .     
\end{equation}  
Equating the cubic terms  in  (\ref{HJ-cubic}), and symmetrizing the second cubic term, produces 
\begin{eqnarray*}   
& &  \beta \sum_{k_1,k_2,k_3}  N_{k_1k_2k_3}  (  M_{k_1} +   M_{k_2} + M_{k_3}  )                        
       \lambda_{k_1}\lambda_{k_2}\lambda_{k_3}   \\ \nonumber      
  & & \;\;\;\;\;\;\; -  \frac{i}{6 \beta} \sum_{k_1+k_2+k_3=0} ( k_1 M_{k_1} +  k_2 M_{k_2} + k_3 M_{k_3} )                           
      \lambda_{k_1}\lambda_{k_2}\lambda_{k_2}   \; = \, 0   \, .   
\end{eqnarray*}
It follows that
\[
 N_{k_1k_2k_3}  = 0 \, , \;\;\;\;\;\;\;\; \mbox{ unless } \;  k_1+k_2+k_3=0 \, , 
\]
and for index triples with $k_1+k_2+k_3=0$,  
\begin{eqnarray}   \label{Ns}   
 N_{k_1k_2k_3}  &=&  \frac{i}{6 \beta^2}  
     \frac{k_1 M_{k_1} +  k_2 M_{k_2} + k_3 M_{k_3}} { M_{k_1} +  M_{k_2} + M_{k_3 } } 
          \\ \nonumber  
      &=&      \frac{i}{6 \beta^2}  
     \frac{k_1 |k_1| +  k_2 |k_2| + k_3 |k_3|} { |k_1| +  |k_2| + |k_3| }    \,  ,
\end{eqnarray}  
in light of (\ref{Mk}).    We note that  $ N_{k_1k_2k_3}$ is symmetric under permutation
of its indices, and that $  N_{-k_1,-k_2,-k_3} =  N_{k_1,k_2,k_3}^*  $.     
 Thus, the coefficients of the approximate solution  (\ref{taylor})
are determined.

This explicit expression for the value function $v(\lambda)$ furnishes 
the gradient term in the closed reduced equation (\ref{closed-reduced-2}); 
namely,  for $k=1, \ldots , m$,    
\begin{eqnarray*}   \label{grad-cubic}
-  \frac{\d v}{ \d \lambda_k^*}    & = &  
      - M_k \lambda_k \, -  \frac{\d }{ \d \lambda_k^*}
      \sum_{k_1+k_2+k_3=0}  N_{k_1k_2k_3}  \, \lambda_{k_1}\lambda_{k_2}\lambda_{k_3} 
         \\ \nonumber
         &=&   - M_k \lambda_k \, - 3  \sum_{k_1+k_2=k}   N_{k_1,k_2, -k}  \, \lambda_{k_1}\lambda_{k_2} 
           \\  \nonumber    
         &=&   - M_k \lambda_k \, - \frac{i}{2 \beta^2}  \sum_{k_1+k_2=k} 
                             \frac{k_1 |k_1| +  k_2 |k_2|  - k^2} { |k_1| +  |k_2| +k }    \,       
                                             \lambda_{k_1}\lambda_{k_2}   \,    ,             
\end{eqnarray*}  
using the symmetry of $N_{k_1k_2k_3}$ in the second equality.

The closed reduced equation, accurate to second order in amplitude, for the resolved vector 
$a = ( \, \la z_1 \, | \, \rhotilde \ra, \, \ldots ,  \,   \la z_m \, | \, \rhotilde \ra \, )$ 
is thus found to be   
\begin{equation}   \label{closed-reduced-3}   
\frac{ d \ahat_k}{d t} \, + \, \frac{i k}{2 } \sum_{k_1 + k_2 = k}  [ \, 1+ \omega(k_1, k_2) \, ] 
\, \ahat_{k_1}   \ahat_{k_2}      \; = \; - \sqrt{\frac{\gamma}{\beta}} \,  |k| \, \ahat_k   \, ,  
\end{equation}    
in which 
\[
\omega(k_1, k_2) \,  =  \, 
\frac{ k_1 |k_1|+ k_2 |k_2| - (k_1+k_2)^2} { (k_1+k_2)( |k_1| + |k_2| + k_1 + k_2) }  \, .
\]
As in the TBH dynamics itself, the indices $k_1$ and $k_2$ in the convolution-like sum run over
$\{ \pm1, \ldots , \pm m  \}$.   The effective dynamics (\ref{closed-reduced-3}) for the $m$ resolved
modes is our main result for the stationary version of the best-fit closure scheme.    The dissipative
terms coincide exactly with those obtained above in the near-equilibrium, linear-response regime.  
The quadratic interaction terms in (\ref{closed-reduced-3}) have the same form as the
corresponding terms in the TBH dynamics (\ref{tbh}), but they are modified by addition of the factors 
$\omega(k_1, k_2)$.      Thus, the best-fit closure procedure produces governing equations
for  the mean resolved modes which capture both the linear dissipation and the
modified nonlinear interactions of the resolved modes
which result from couplings between the resolved and unresolved modes.      

The dissipative structure of (\ref{closed-reduced-3}) is novel in the sense that it 
does not correspond to a standard diffusion, for which the decay of mode $k$
scales with $k^2$.  Rather the dissipation for the closure of the TBH dynamics
is a fractional diffusion, for which  the decay of mode $k$ scales with $|k|$.     
As mentioned above, this prediction of our closure theory is robustly observed in numerical simulations,
including those presented in Sections 7 and 8.   

The effects of the modification of the TBH mode interactions are more subtle.   
The factors $\omega(k_1,k_2) $ can be either positive or negative depending
on the signs of $k_1$ and $k_2$.   Specifically, 
\[
- \frac{1}{4} \, \le \omega(k_1, k_2)  \, = \,  \frac{- k_1 k_2}{(k_1 + k_2)^2} \,  < \,  0, 
   \;\;\;\;\;\;\;\;   \mbox{ when } \;\;    k_1> 0 , \,  k_2 > 0 \, ,  
\]
\[
0  \, < \, \omega(k_1, k_2)  \, =  \, \frac{|k_2|}{k_1} \,  < \,  1 
\;\;\;\;\;\;\;\; \;\;\;  \mbox{ when } \;\;    k_1 > 0,  \, k_2 < 0 \, ,   
\]
and symmetrically when $k_1<0, \, k_2 >0$.   
Accordingly, the low-to-high mode transfers for the closed reduced equation 
are weaker than for the TBH equation itself, while the high-to-low mode transfers are stronger.     
Also, there is an asymmetry between the
weakening and the strengthening of the mode transfers:  The downscale (low-to-high mode)
transfers are weakened by factors of at most 25\%, while the upscale 
(high-to-low mode) transfers are strengthened by factors that can approach 100\%.

\section{Nonstationary formulation of the closure}   

The stationary formulation of the best-fit closure scheme developed in the preceding section 
produces a time-independent value function, $v(\lambda)$, which determines the
irreversible component of the flux of the estimated mean resolved vector $\ahat$ according to
(\ref{closed-reduced-1}).    The entropy production along an extremal $\lambdahat(t) $
for the stationary optimization principle (\ref{value-fn})  is given by 
\[
\frac{d \hat{s}}{dt}  \, = \,  \;   \sum_{k=-m}^m 
        \lambdahat_k \, \frac{\d v}{\d \lambda_k }(\lambdahat)       \, . 
\]
It follows that throughout the subdomain over which the value function is convex,
which includes a neighborhood  of equilibrium, 
\[
\frac{d \hat{s}}{dt} \, \ge  \, v(\lambdahat) \, > 0 \, , 
\]
since, by convexity,  $0=v(0)  \ge  v(\lambda) - \sum_{-m}^m \lambda_k\d v / \d \lambda_k $.    
Conceptually, this inequality is entirely satisfactory:  the entropy production
is bounded below by the value function, which quantifies the optimal information
loss in the closure and depends only on the statistical state $\lambdahat$.  
But, when a quasi-equilibrium density $\rhotilde(\lambda^0)$  is specified as
the initial state at $t=0$, the entropy production of the path of quasi-equilibrium states
emanating from $\rhotilde(\lambda^0)$ and following the exact mean values
$a(t)$  tends to zero as $t \rightarrow 0+$.  
In this situation the entropy production, and the irreversible component
of the flux, vanish initially and there is a ``plateau'' time over which they grow
to their stationary values \cite{Tur12}.     Since our numerical experiments in 
Sections 7 and 8 make use of such initial conditions, it is necessary to modify the formulation of
the defining optimization principle to include this plateau effect.    We refer to this modification
as the nonstationary formulation, in that the value function $v =v(\lambda,t)$ becomes
time dependent, and $v(\lambda,0)=0$ identically.      

The nonstationary optimization principle has the value function   
\begin{equation}    \label{value-fn-non}
v(\lambda^1,t_1) \, = \,   \min_{\lambda(t_1)=\lambda^1} \; \int_{0}^{t_1}  
        \L(\lambda, - \lambdadot)  \, dt \,  , 
 \end{equation}   
in which the admissible paths 
$\lambda(t), \; 0 \le t \le t_1 \, $, in the configuration space of the
statistical model are constrained to terminate  at $\lambda^1 \in \C^n$ at time $t_1 \ge 0$,
while $\lambda(0)$ is unconstrained.   The integrand in (\ref{value-fn-non}) is modified 
to account for time-reversal.   The admissible paths may be viewed as evolving in reversed time,
$\tau = t_1 -t$, starting from $\lambda^1$ at $\tau=0$.  The value function at $t=t_1$ then quantifies
the optimal lack-of-fit of time-reversed paths connecting the current state $\lambda^1$ 
to some (unspecified) initial state $\lambda(0)$  corresponding to    
some (quasi-equilibrium) trial density at $t=0$.  A fuller justification is given in \cite{Tur12}. 
 
In terms of this value function (\ref{value-fn-non}), closure is achieved at each time $t$ 
 by setting (now with $(\lambda,t)$ replacing $(\lambda^1,t_1$) )
\begin{equation}  
\hat{\mu} (t) \, = \, - \frac{ \d v}{ \d \lambda^* } (\lambda, t)   \, .
\end{equation}   
This relation is the nonstationary analogue to (\ref{conjugate-relation}), and it      
determines the reduced dynamics in the same way that
(\ref{closed-reduced-1}) follows from (\ref{conjugate-relation}).   
The resulting  equations for the mean resolved variables are
the nonstationary version of (\ref{closed-reduced-2}); namely,
\begin{equation}   \label{closed-reduced-non}  
 \frac{d \ahat_k}{dt} + \frac{ik}{2} \sum_{k_1 +k_2 = k} \ahat_{k_1} \ahat_{k_2} 
    \, = \,   -  \frac{\d v}{\d \lambda_k^*}(\beta \ahat, t)    \, .  
\end{equation}    
The value function (\ref{value-fn-non}) is the unique solution of the initial value
problem 
\begin{equation}  \label{HJ-non}
 \frac{\d v}{\d t}  \, + \,  \H \left( \lambda, \, - \frac{\d v}{\d \lambda^*} \right) \, = \, 0 \, , \;\;
    \mbox{ for } \; t>0,  
    \;\;\;\;\; \mbox{ with } \;   v(\lambda,0)=0 \, ,  
\end{equation}
which is a time-reversed Hamilton-Jacobi equation;  that is,  
$\H(\lambda, - \mu)$ is the Legendre transform of $\L(\lambda, - \lambdadot)$.        

Since $\H(\lambda,\mu)$ is positive-definite in $\mu$, the solution $v(\lambda,t)$
of (\ref{HJ-non}) exists for all time $t>0$, and    as $t \rightarrow +\infty$,  
$v(\lambda,t) \rightarrow v(\lambda)$, the stationary value function.    
Thus, the nonstationary best-fit closure is a natural generalization 
of the stationary closure that straightforwardly includes an intrinsic plateau effect.     

As in Section 5,  the closure that holds in the near equilibrium
regime is calculable by elementary methods.   
In the defining nonstationary 
optimization principle (\ref{value-fn-non}) the extremal path $\lambdahat(t)$ satisfies
$d \lambdahat/dt(0) =0$, because $\lambda(0)$ is free;
solving the Euler-Lagrange equations with this initial condition gives 
\[
\lambdahat_k(t) = \lambdahat_k(t_1) \frac{ \cosh \sqrt{\frac{\gamma}{\beta}} |k| t }
                                                                  { \cosh \sqrt{\frac{\gamma}{\beta}} |k| t_1 }  \, ,
\] 
an extremal that decays exponentially in reversed time.    
The near-equilibrium, nonstationary, closed reduced equations  are accordingly    
\begin{equation}  \label{simple-cr-non}
\frac{ d \ahat_k}{ dt}  \, = \, 
  -  \sqrt{\frac{\gamma}{\beta}} \, |k| \,   \tanh \left(  \sqrt{\frac{\gamma}{\beta}} |k| t  \right)  \ahat_k    \, .
\end{equation} 
These equations clearly approach the corresponding stationary equations (\ref{simple-cr})
asymptotically as $t \rightarrow +\infty$.    
 The value
function evaluated on this family of extremals is calculated to be
\[
v(\lambda,t)  \, = \,    \sqrt{\frac{\gamma}{\beta^3}} \, \sum_{k=1}^m k \, |\lambda_k|^2 \, 
                                      \tanh \left(  \sqrt{\frac{\gamma}{\beta}} |k| t  \right)   
\]

Proceeding to the quadratically nonlinear corrections for the nonstationary closed reduced
equations, we consider a Taylor expansion (\ref{taylor}) for $v(\lambda, t)$, in which
now the coefficients are time-dependent,  $M_k(t), \; N_{k_1,k_2, k_3} (t)$, and
$M_k(0)=0, \; N_{k_1,k_2, k_3} (0)=0$.     The analysis of Section 5 then follows along
the same lines as in the stationary version, except that now $M_k$ and $N_{k_1,k_2, k_3}$
satisfy ordinary differential equations rather than algebraic equations.    Specifically, 
the quadratic and cubic terms in the expansion of the time-reversed Hamilton-Jacobi equation 
(\ref{HJ-non}) yield the ODEs
\begin{equation}   \label{m-k-non}
\frac{ dM_k}{dt}  \, + \, \beta M_k^2 \, = \,  \frac{\gamma}{\beta^2} k^2   \, ,
\end{equation} 
\begin{equation}  \label{n-kkk-non}
 \frac{ d N_{k_1k_2k_3}}{dt} \, + \, \beta  (  M_{k_1} +   M_{k_2} + M_{k_3}  ) \, N_{k_1k_2k_3}                  
         \; = \,   \frac{i}{6 \beta}  ( k_1 M_{k_1} +  k_2 M_{k_2} + k_3 M_{k_3} )     \, .  
\end{equation}  
The solution of the Riccati differential equation (\ref{m-k-non}) with $M_k(0)=0$ is 
\[
 M_k(t) \, = \, \sqrt{\frac{\gamma}{\beta^3}} \, |k| \,   \tanh \left(  \sqrt{\frac{\gamma}{\beta}} |k| t  \right)   \, ;
\]
this result produces the linear terms in (\ref{closed-reduced-non}), which are anticipated
in (\ref{simple-cr-non}).    
Inclusion of the quadratic terms in (\ref{closed-reduced-non}) associated with the 
coefficient functions $N_{k_1k_2k_3}(t) $ produces the nonstationary version of the 
closed reduced equations (\ref{closed-reduced-2}).   Namely,   for $k=1, \ldots, m$, 
\begin{equation}    \label{closed-reduced-non-2}
\frac{ d \ahat_k}{d t} \, + \, \frac{i k}{2 } \sum_{k_1 + k_2 = k}  [ \, 1+ \Omega(k_1, k_2,t) \, ] 
\, \ahat_{k_1}   \ahat_{k_2}      \; = \; 
  - \sqrt{\frac{\gamma}{\beta}} \, |k| \, \tanh \left( \sqrt{\frac{\gamma}{\beta}} \, |k| \, t \right)  \ahat_k   \, ,  
\end{equation}  
where the functions $\Omega(k_1,k_2,t)$ solve the ODEs
\begin{eqnarray}  \label{omega}
 \frac{ d \, \Omega(k_1,k_2,t)}{dt} & + & 
         \beta \left[ \,  M_{k_1}(t) + M_{k_2}(t) +M_{k}(t)  \,  \right] \, \Omega(k_1,k_2,t)  \\ \nonumber 
         & = & \frac{\beta}{k} \, \left[ \, k_1M_{k_1}(t) + k_2 M_{k_2}(t) - k M_{k}(t) \,   \right]  
       \hspace{1cm}  ( \,  k_1+k_2 = k \, ) \,  , 
\end{eqnarray} 
along with homogeneous initial conditions $\Omega(k_1,k_2,0)=0$.  These ODEs follow
immediately from (\ref{n-kkk-non}) by setting 
 $ \; \Omega(k_1,k_2,t) = ( 6 \beta^2/ik)N_{k_1,k_2,-k}(t) \; $  for $k_1+k_2 = k >0$.

Even though the equations for the factors $\Omega(k_1,k_2,t) $ are inhomogeneous linear ODEs,
their coefficients and source terms are time-dependent, and hence they are not solvable analytically.
A simple approximation, which may be helpful conceptually, is to replace the functions $M_k(t)$
in (\ref{omega}) by their saturated values, $\lim_{t \rightarrow + \infty} M_k(t)$.    Then the time-dependent
factors  $\Omega(k_1,k_2,t) $ are simply related to the time-independent factors  $\omega(k_1,k_2) $
in (\ref{closed-reduced-3});  namely,  
\[
\Omega(k_1, k_2,t) \,  \approx  \, \omega(k_1, k_2) \cdot 
     \left\{ \, 1 - \exp \left( -  \sqrt{ \frac{\gamma}{\beta} } 
             \left[ \, |k_1|+|k_2|+k_1+k_2   \right] \, t \, \right) \right\}  \, . 
\]
It is transparent from these approximate expressions that the modifications of the nonlinear
interactions in the nonstationary formulation saturate to the stationary formulation with
saturation rates  that are mode dependent.   

The nonstationary closed reduced equations  (\ref{closed-reduced-non-2}) constitute our
final result of the best-fit closure analysis.    We emphasize that this coarse-graining of
the  TBH dynamics depends only on a single closure parameter, $\gamma$.  
By adjusting $\gamma$ the best-fit theory endeavors to model the linear dissipation,
the modifications of the nonlinear interactions, and the plateau effect, all of which are mode dependent.  
The quantitative validation of this model via numerical experiments is described in next section.

\section{Numerical model and experimental setup}

\subsection{Direct numerical simulations (DNS)}
To ensure that the equilibrium distribution of the TBH model is close to the Gibbs
distribution (\ref{gibbs}),  we choose $n=50$ complex modes.  This particular value has been tested extensively numerically in the literature \cite{mati01}.
To obtain a reduced model that is much smaller but still
resonably complicated, we choose $m =5$, retaining the 5 complex modes of smallest
wavenumber, $k=1,\ldots,5$.    Thus, our reduction is from 100 independent real modes to
10 real modes.   

For the prognostic numerical integration of the Fourier modes
$z_{k}(t)$, $k=1, \ldots ,50$, we use a fourth order Runge-Kutta timestepping method.
The nonlinear advection terms are calculated by evaluating the Fast
Fourier transform of $P_{n}\left(\frac{1}{2}u_{n}^{2}\right)$ and
then multiplying by $ik$;  that is, a pseudo-spectral method is utilized
to integrate equations (\ref{tbh-xform}) and (\ref{tbh}). This algorithm
has been found by \cite{mati01} to respect the conservation of $E$ to
a high accuracy; throughout the time integrations reported
here we observe conservation to an accuracy of around $10^{-5}$,
 the expected value of $E$ being $10.0$.

Ensembles are constructed by sampling initial conditions from a particular
trial distribution of the form (\ref{trial-densities}). 
The $100$ real modes under these distributions are Gaussian with
variances all equal to $1/2\beta$.   A value of $\beta=5.0$ is adopted 
for the inverse temperature, 
 corresponding to a standard deviation of $1/\sqrt{10}$ 
for the real and imaginary parts of all $z_{k}$.
The means of the resolved complex modes, $z_1, \ldots, z_5$, for
the initial trial distribution 
trial distributions are derived as follows: From a long and hence
equilibrated integration of the numerical model we draw a  vector 
$b_{k}\equiv z_{k}(T)$, $k=1, \ldots, 5$; the initial means, $a_{k}(0)$,
are then specified by multiplying  $b_{k}$  by a fixed factor $r_{dev}$.   
The ensemble members for the initial distribution are drawn by sampling
the resulting Gaussian distribution. This approach allows us 
 to test the sensitivity of results on the magnitude of the initial
deviation from equilibrium. We examine the two cases,  $r_{dev}=1/\sqrt{10}$
and $r_{dev}=\sqrt{10}$,  and refer to them below as the ``close to'' and ``significantly removed from''
equilibrium experiments, respectively. We comment below on other choices for $r_{dev}$. 
The results reported here are
found to be qualitatively similar when different choices for the $b_{k}$
are adopted. We thus report results from only one particular set
of randomly generated $b_{k}$. 

Since the focus of this study is on the time evolution of the resolved 
mean variables,  $a_{k}(t)=\left\langle z_{k}| \rhotilde(\lambda(t)) \right\rangle $,
we construct ensembles of sufficient size so that this first
moment is statistically steady.   Emprically this has been determined to
be more than met by samples of size $10^{6}$, which we adopt here.   
As an objective test of the ensemble size issue, we have examined subsample
results with size $10^{5}$ and have noted only very small differences
to the results.   As we shall discuss further
below, the ensemble means $a_{k}(t)\rightarrow0 $ for sufficiently
large $t$, and in fact the time scale for this first moment equilibration
varies inversely with the wavenumber $k$.   The results
we report are integrated until $t=1.5$, which ensure all the complex
modes except the first  have very clearly equilibrated first moments;
the first complex mode ($k=1$) has nearly but not completely equilibrated
after this integration time.

\subsection{Closure model computations}

The appropriate closure equation for comparison with the DNS results
is given by equations (\ref{closed-reduced-non-2}) and (\ref{omega}) in Section 6. 
There are two important features of these two sets of equations. The reduced equation
(\ref{closed-reduced-non-2}) controlling the first moment evolution is similar in form to
the original DNS model but with modified advection coefficients and
with wavenumber-dependent damping.   To solve it we modify the original TBH integrator
by truncating to $m$ complex modes and revert the  calculation of the advection term 
to a purely spectral calculation that uses the time-dependent factors $\Omega$.
These factors, which modify the advection coefficients, are determined by the decoupled,
forced and damped, linear ODEs (\ref{omega}), and hence they are computed 
by a simple two time-step in which the damping and forcing are evaluated on the backward timestep.   
The wavenumber-dependent damping in (\ref{closed-reduced-non-2}) is evaluated at the 
first time step of the fourth-order Runge-Kutta scheme for  advection.  

The closure equations have one free parameter $\gamma$ which we determine
as follows: The total squared difference between the $2m$ closure
modes and the first $2m$ variables of the DNS at each time step (which is .0015) is
designated the error function for the fitting execise. It is evaluated
numerically for a large number of choices for $\gamma$ by running
both equations (\ref{closed-reduced-non-2}) and (\ref{omega}) many times, a very cheap exercise.
It is found to be always convex with a unique minimum which is determined
by a manual convergence technique.

\subsection{Modified closure procedure}

The time scale separation between the retained modes and the discarded
modes is not very sharp, a common feature for turbulent systems. One might
expect intutively for this situation to adversely affect the closure
performance. In order to evaluate this possibility we extend the closure model 
from $2m$ resolved modes to $4m$ resolved modes
and regard the additional $2m$ modes as inserting a buffer between
the fast and slow parts of the system.  The extended closure model has
$2m$ slow modes, $2m$ intermediate modes, and 
$16m$ ``ignored'' modes with the highest wavenumbers and the fastest time
scales, which are regarded as a ``heat bath.'' The closure model
is run as before but the buffer modes are initialized at mean zero, while
the original resolved $2m$ modes are initialized by  $r_{dev}b_{k}$ as
in the previous subsection.  

\section{Numerical results}

\begin{figure}
        \begin{subfigure}[b]{0.67\textwidth}
                \centering
                \includegraphics[width=\textwidth]{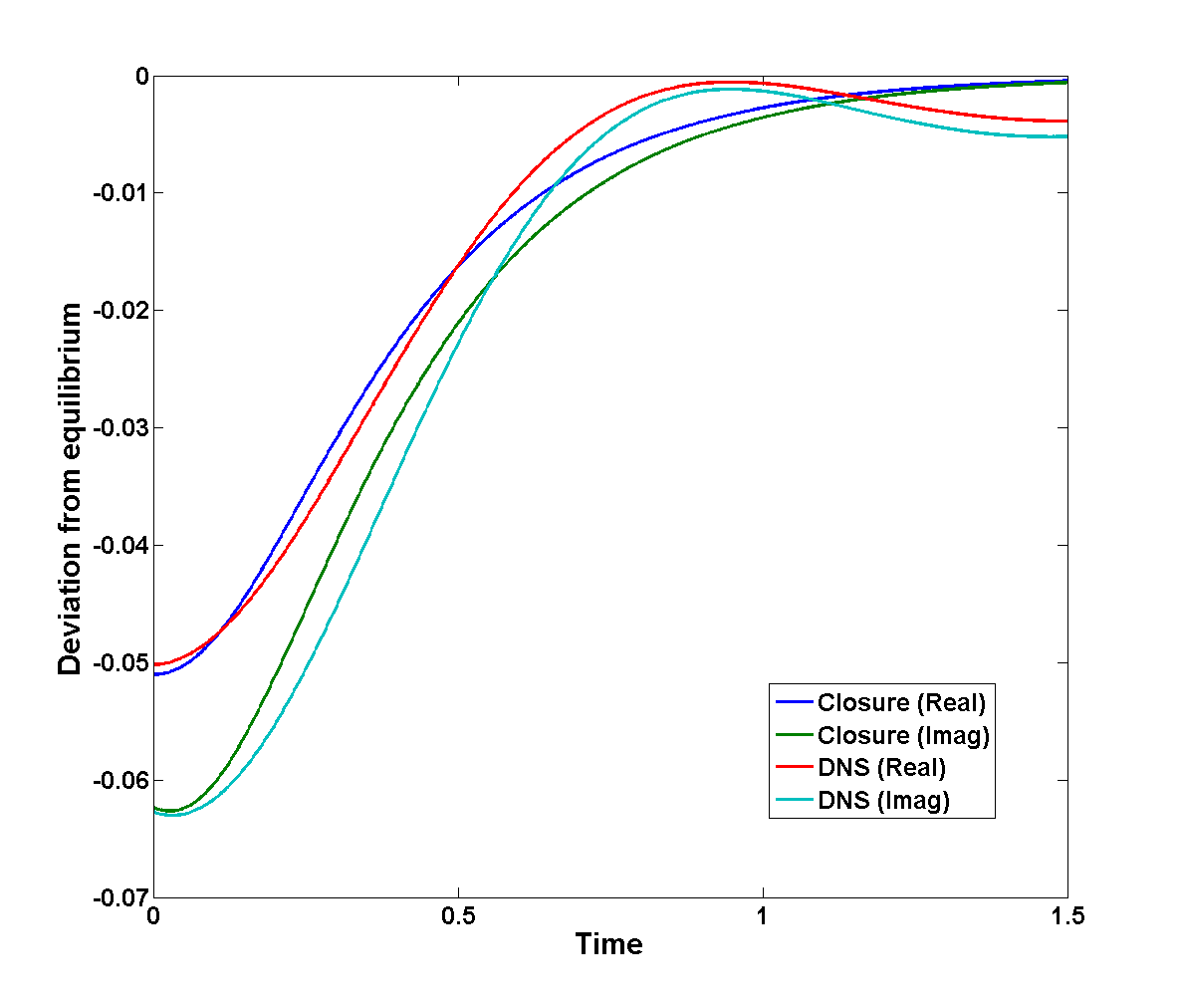}                
                \caption{Complex Mode 1}
                \label{fig:1a}
        \end{subfigure}

        \begin{subfigure}[b]{0.67\textwidth}
                \centering
                \includegraphics[width=\textwidth]{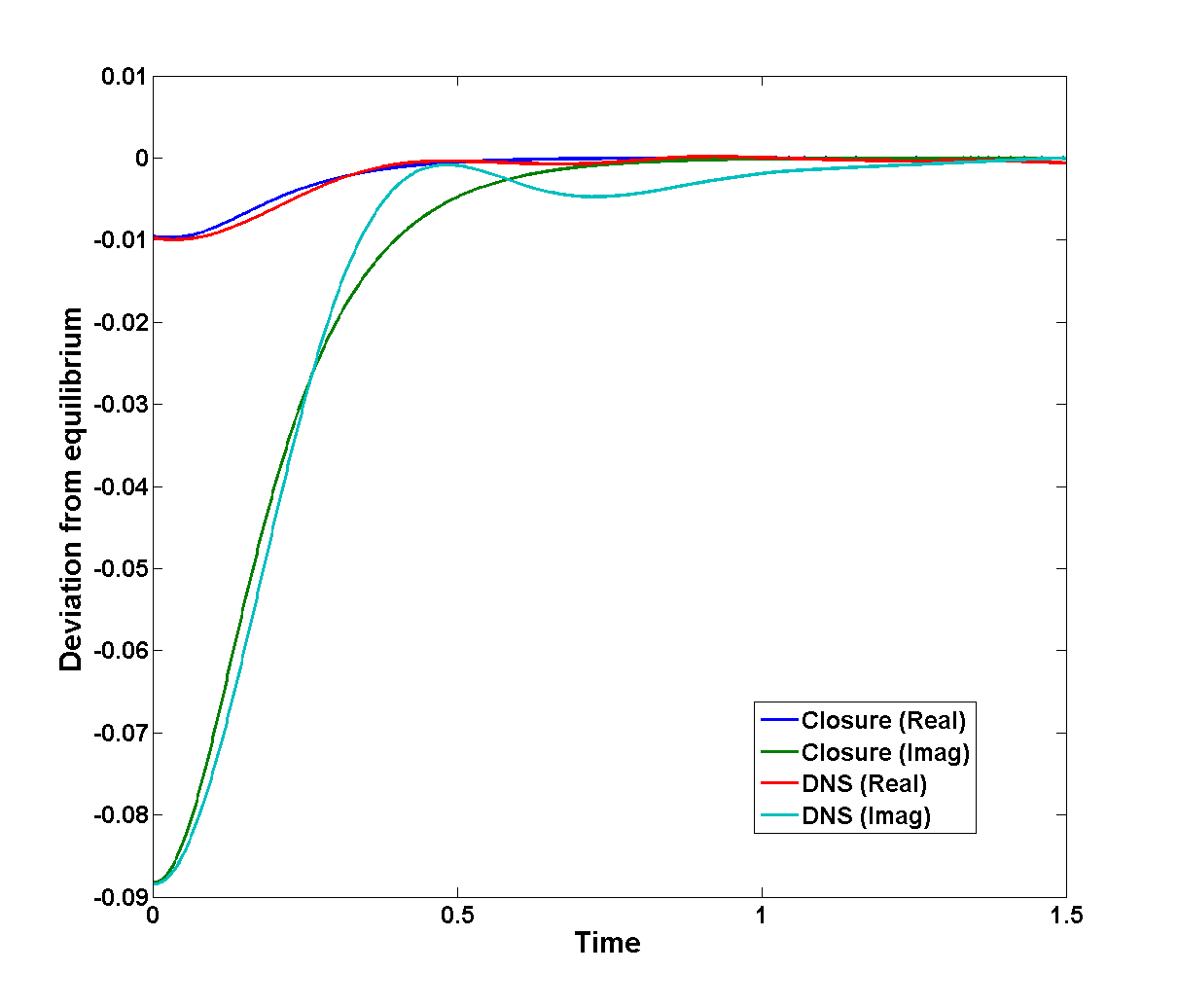}                
                \caption{Complex Mode 2}
                \label{fig:1b}
        \end{subfigure}
\caption{Comparison of the evolution of first moments for the direct numerical simulation (DNS) and the closure model. The initial conditions here
 are close to equilibrium (see text). There are five panels corresponding to the five retained complex modes of lowest wavenumbers. In each panel the real and imaginary part
of each complex mode is plotted for both the DNS and closure. As usual both components of the complex modes have the same wave number.}
\label{Figure1}
\end{figure}
\begin{figure}
     \ContinuedFloat 
     \begin{subfigure}[b]{0.73\textwidth}
                \centering
                \includegraphics[width=\textwidth]{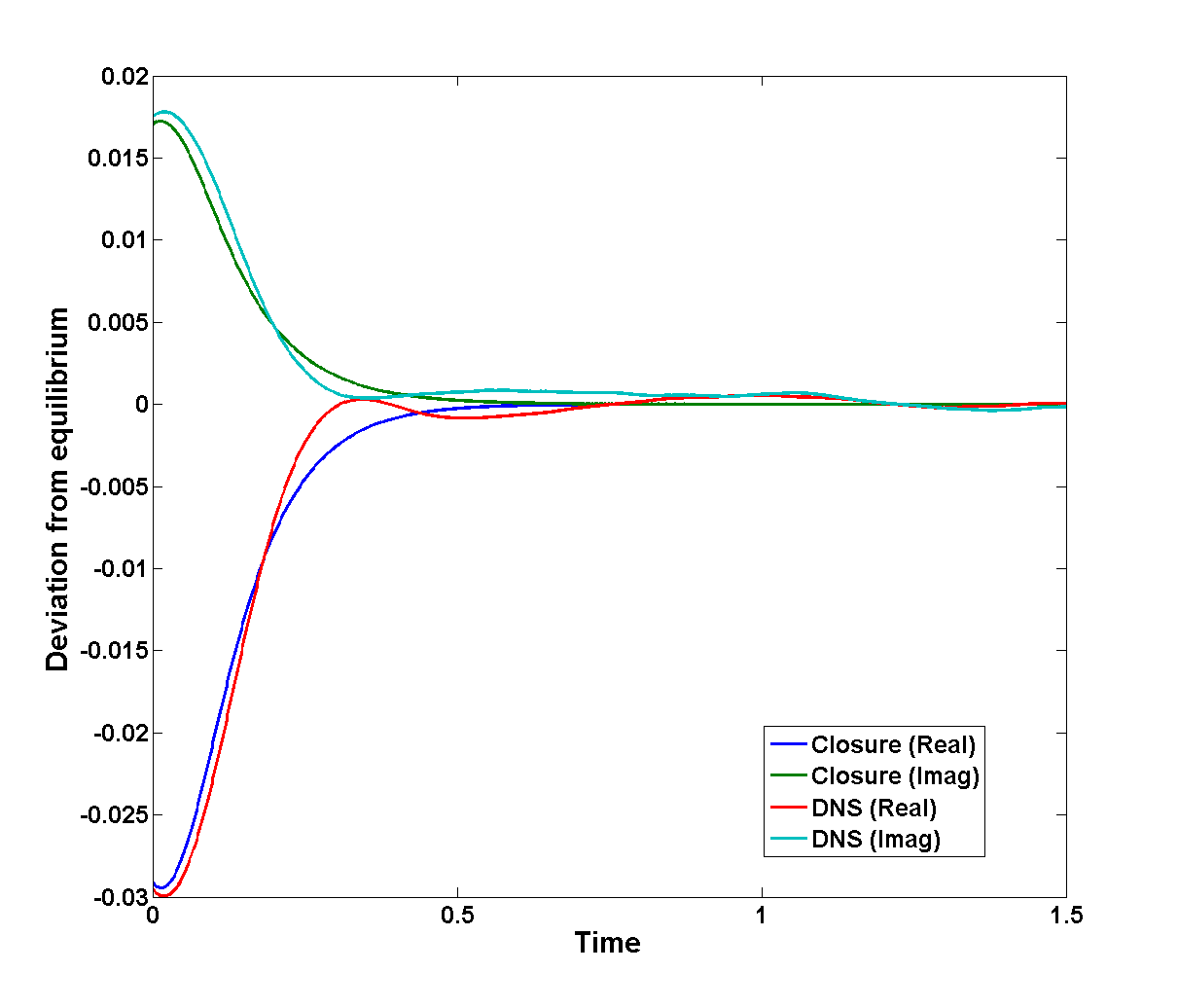}                
                \caption{Complex Mode 3}
                \label{fig:1c}
        \end{subfigure}

        \begin{subfigure}[b]{0.73\textwidth}
                \centering
                \includegraphics[width=\textwidth]{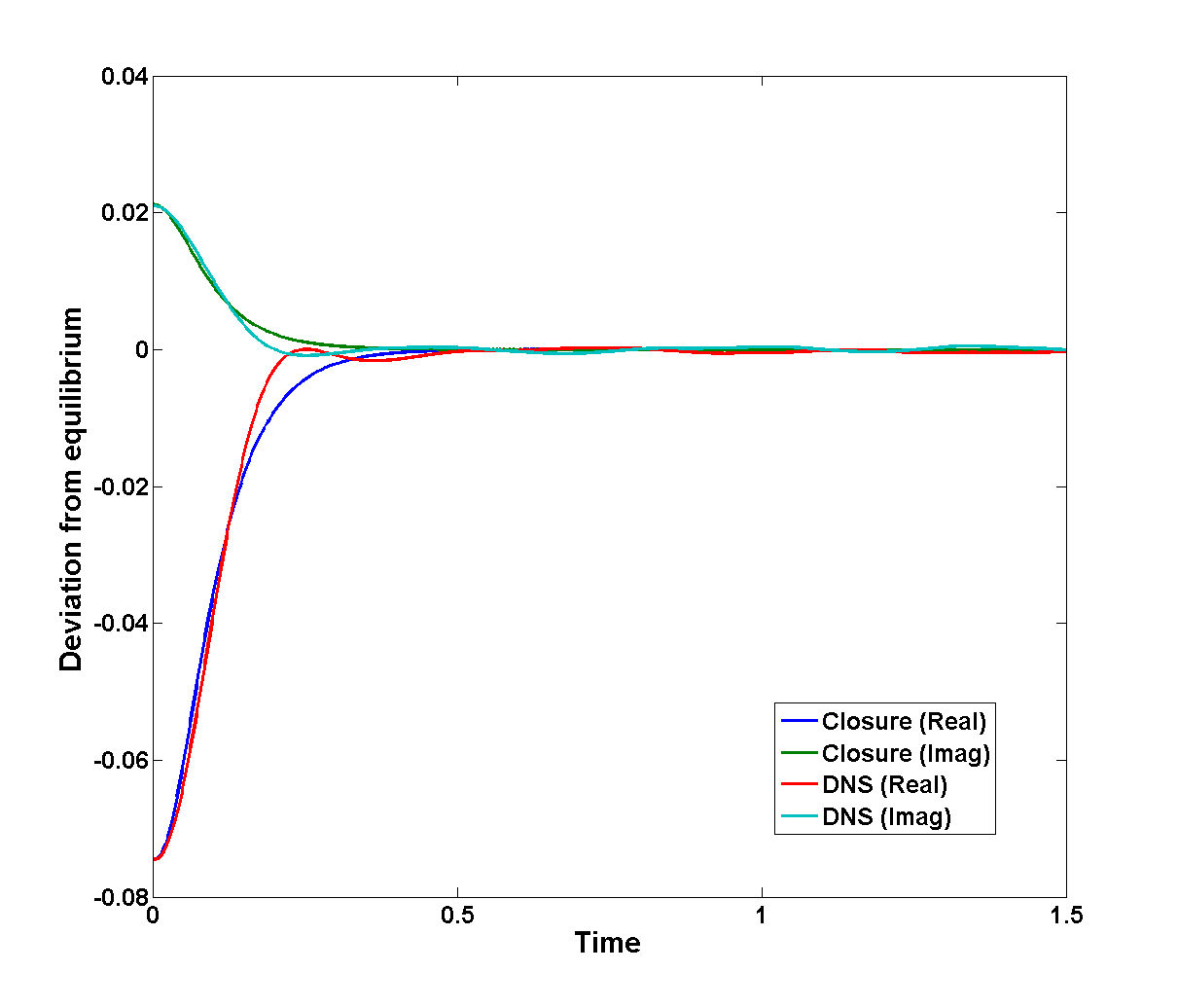}                
                \caption{Complex Mode 4}
                \label{fig:1d}
        \end{subfigure}
\caption{(continued)}
\label{Figure1}
\end{figure}
\begin{figure}
     \ContinuedFloat 
        \begin{subfigure}[b]{0.73\textwidth}
                \centering
                \includegraphics[width=\textwidth]{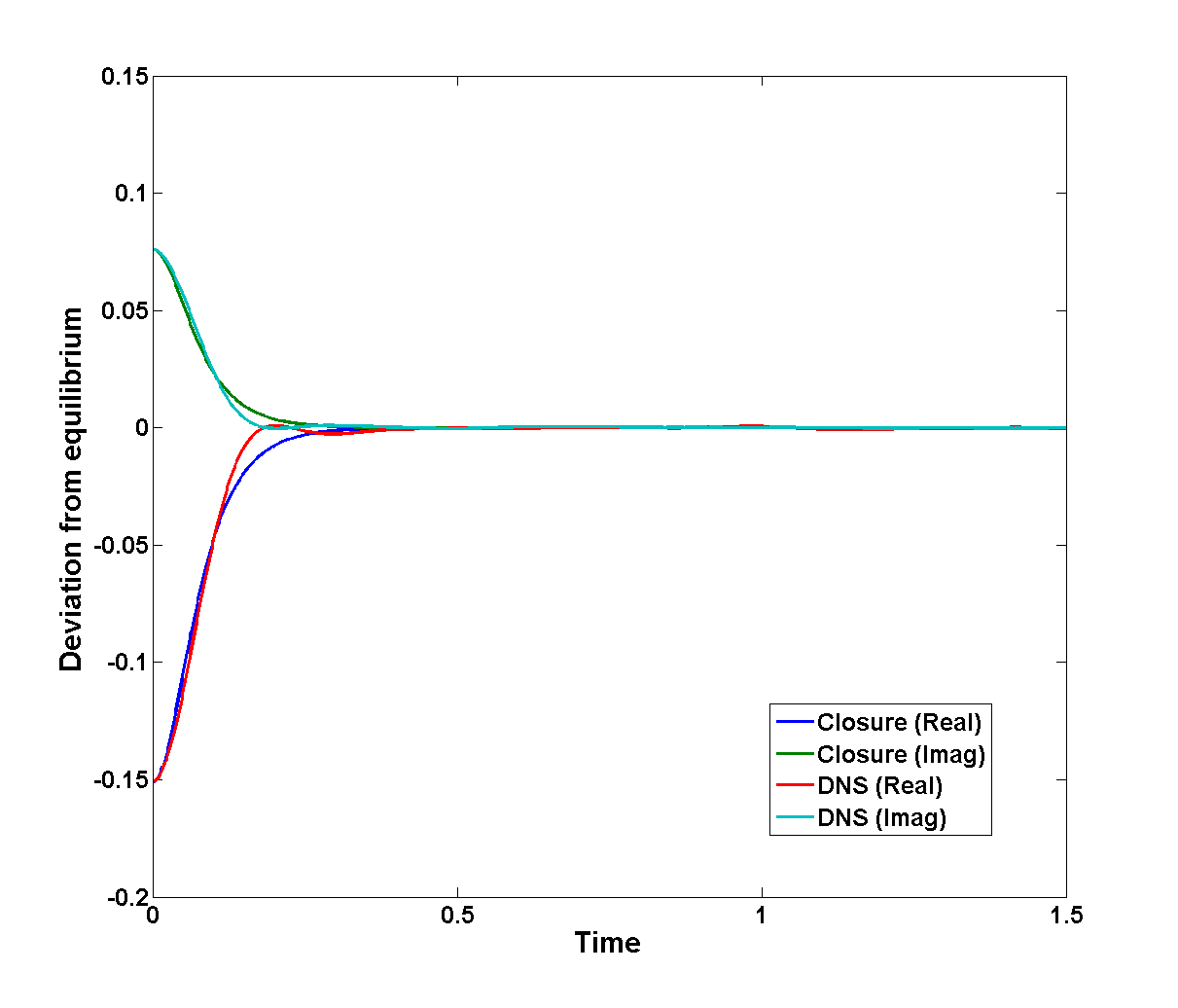}                
                \caption{Complex Mode 5}
                \label{fig:1e}
        \end{subfigure}
\caption{(continued)}
\label{Figure1}
\end{figure}
\subsection{Close to equilibrium case}

A comparison between the  first moments  calculated by the DNS and those calculated
using the closure model with the best choice of $\gamma=64.74$ is displayed
in Figure \ref{Figure1}. Overall the RMS error per mode per timestep
is found to be $0.002035$, which may compared with the amplitude of the
first moments. Each panel shows the behavior of the real
and imaginary parts of $a_{k}(t)$ and all $m=5$ closure modes and
DNS modes are plotted. The performance is quite
good, and in particular two distinctive features of the statistical
dynamics are well simulated qualitatively by the closure:
\begin{enumerate}
\item The DNS equilibration time for each complex mode is very clearly proportional
to wavenumber and excellently reproduced by the closure equations.
This provides strong evidence that the correct dissipation
for reduced models of TBH is given by a fractional diffusion process.
It is notable that exactly this kind of dissipation is the bare minimum
required by the infinite Fourier mode Burgers-Hopf equation to prevent
the appearance of a singular shock in finite time \cite{KNS}.
The optimal value of $\gamma$ reported above corresponds
with an exponential damping time of $0.2779/|k|$, according to
 the simplified analysis at the beginning
of Section 5, 
and this timescale is clearly visible in the approach to equilibrium
of all modes.
\item Each mode exhibits essentially two well-defined regimes during equilibration:
An initial ``plateau'' period and a later exponential decay to equilibrium.
These periods are of comparable duration and are both directly proportional
to wavenumber. Both of these features are predicted by the theoretical
analysis of Section 6 and are readily apparent in Figure \ref{Figure1}, most obviously
for the low wavenumber cases. As the theoretical
analysis shows, the fractional diffusion is ramping up during the plateau period,
suggesting that the evolution of the resolved variables is more under
the control of the initial conditions and to a lesser extent the non-linear
interactions. 
\end{enumerate}
The discrepancy between DNS and closure tends, in general, to increase
with time and is also more apparent in the low wavenumber modes. The
DNS exhibits some oscillatory behaviour as it approaches equilibrium,
which the closure model does not reproduce. These oscillations have
a wavelength inversely proportional to the mode wavenumber.  Nonetheless, a careful
inspection reveals that the closure trajectories tend to ``bisect'' 
these DNS oscillations cleanly in every instance. 

\begin{figure}
        \begin{subfigure}[b]{0.73\textwidth}
                \centering
                \includegraphics[width=\textwidth]{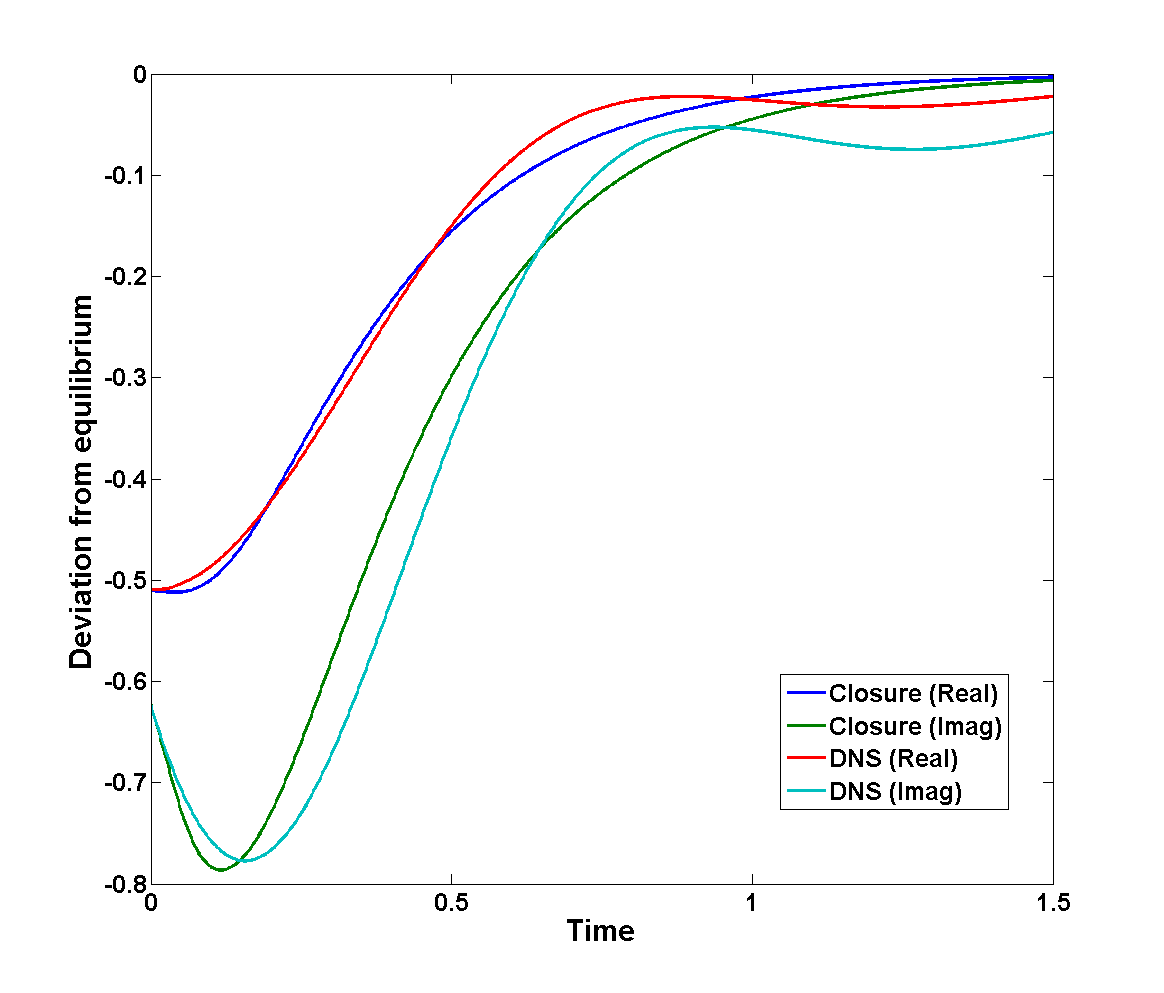}                
                \caption{Complex Mode 1}
                \label{fig:2a}
        \end{subfigure}

        \begin{subfigure}[b]{0.73\textwidth}
                \centering
                \includegraphics[width=\textwidth]{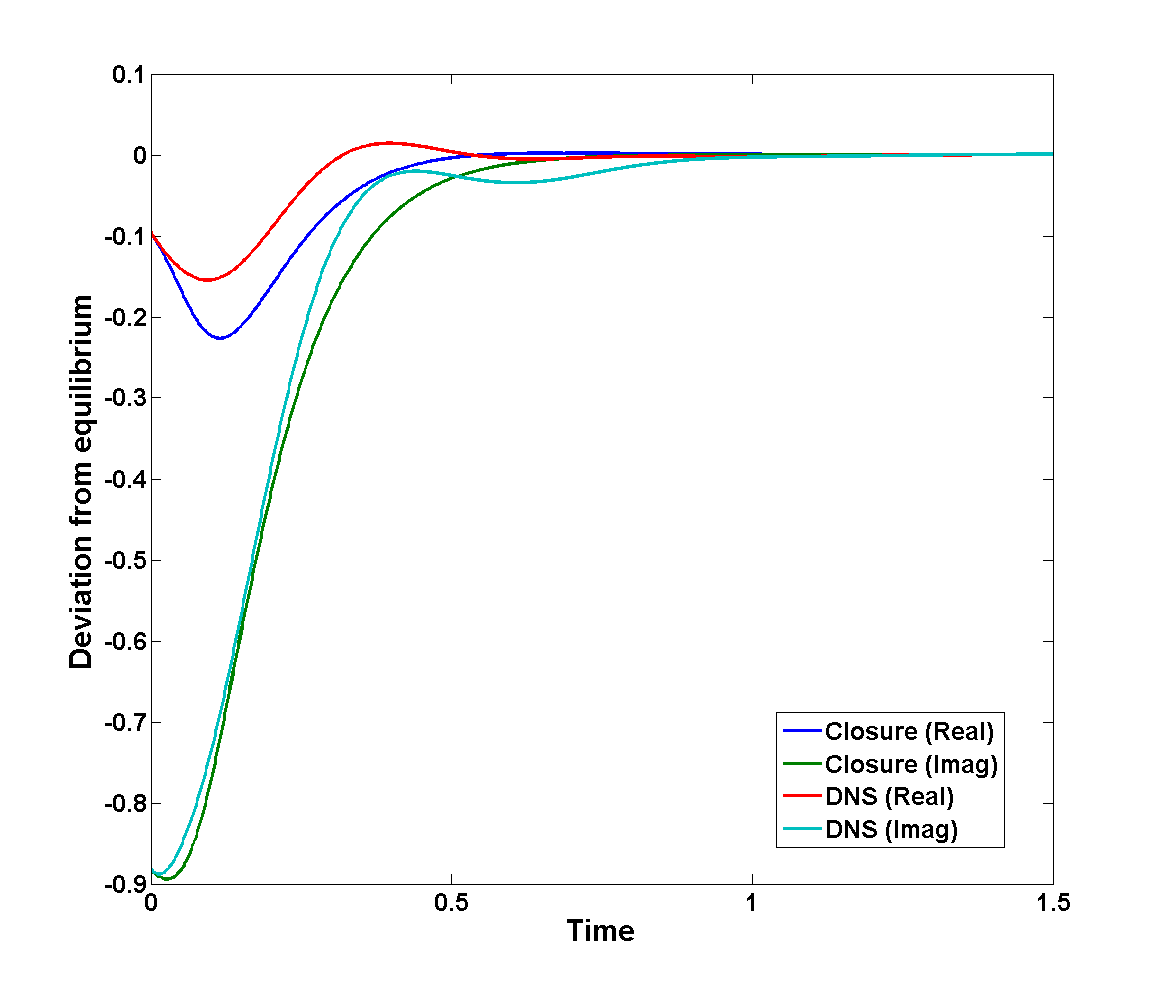}                
                \caption{Complex Mode 2}
                \label{fig:2b}
        \end{subfigure}
   \caption{Same as Figure \ref{Figure1} except the initial conditions here
 are significantly removed from equilibrium (see text)}
\label{Figure2}
\end{figure}

\begin{figure}
     \ContinuedFloat 
        \begin{subfigure}[b]{0.73\textwidth}
                \centering
                \includegraphics[width=\textwidth]{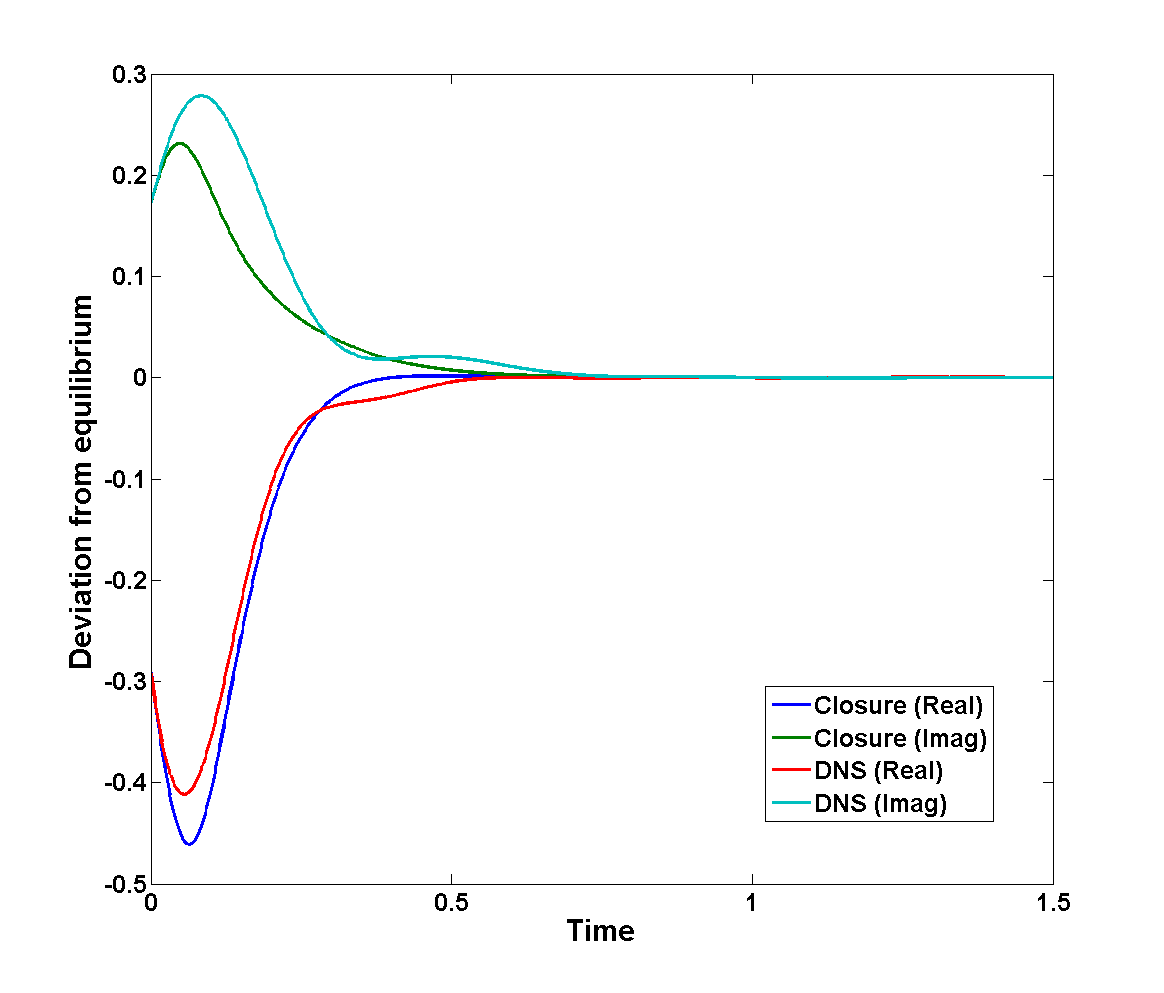}                
                \caption{Complex Mode 3}
                \label{fig:2c}
        \end{subfigure}

        \begin{subfigure}[b]{0.73\textwidth}
                \centering
                \includegraphics[width=\textwidth]{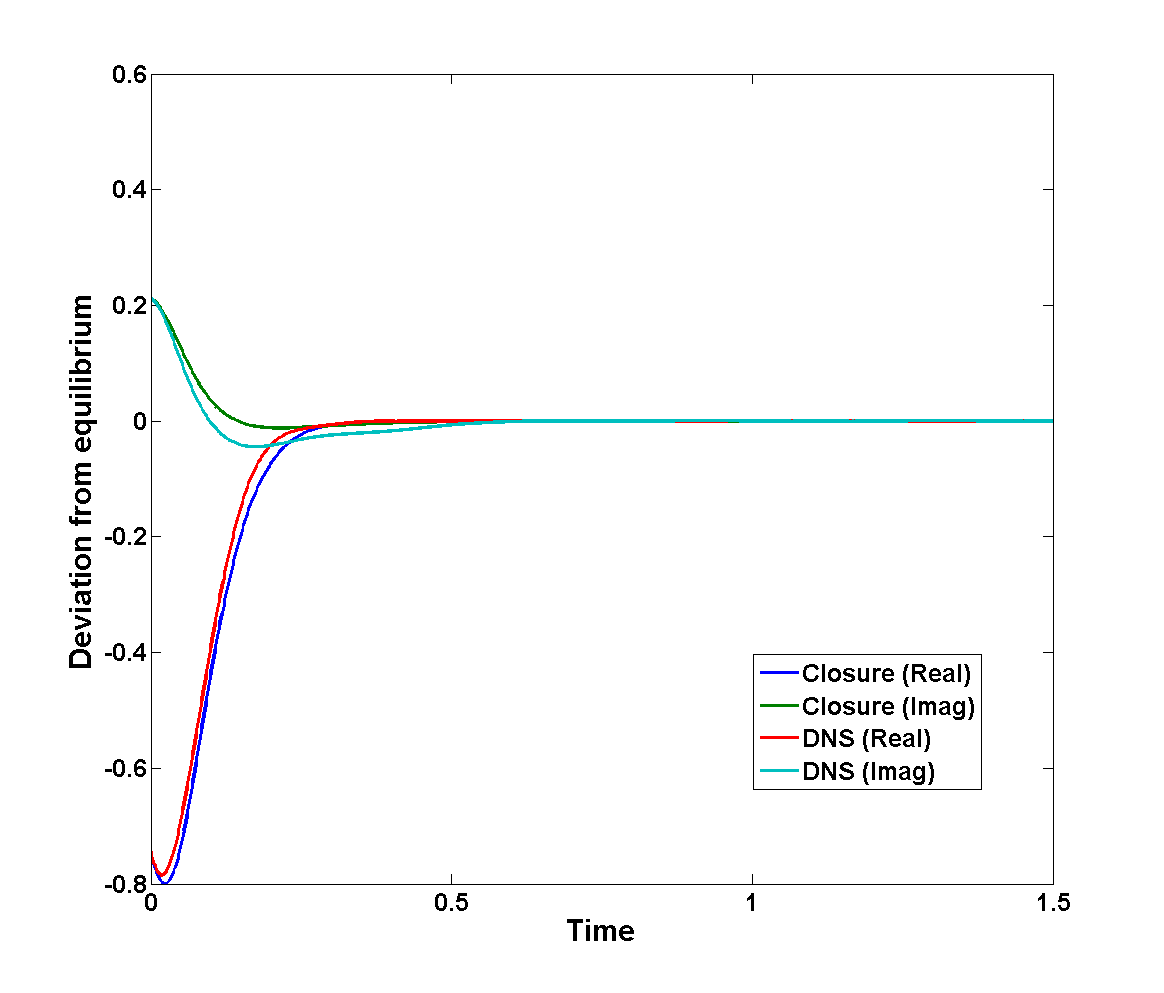}                
                \caption{Complex Mode 4}
                \label{fig:2d}
        \end{subfigure}
\caption{(continued)}
\label{Figure2}
\end{figure}
\begin{figure}
     \ContinuedFloat 

        \begin{subfigure}[b]{0.73\textwidth}
                \centering
                \includegraphics[width=\textwidth]{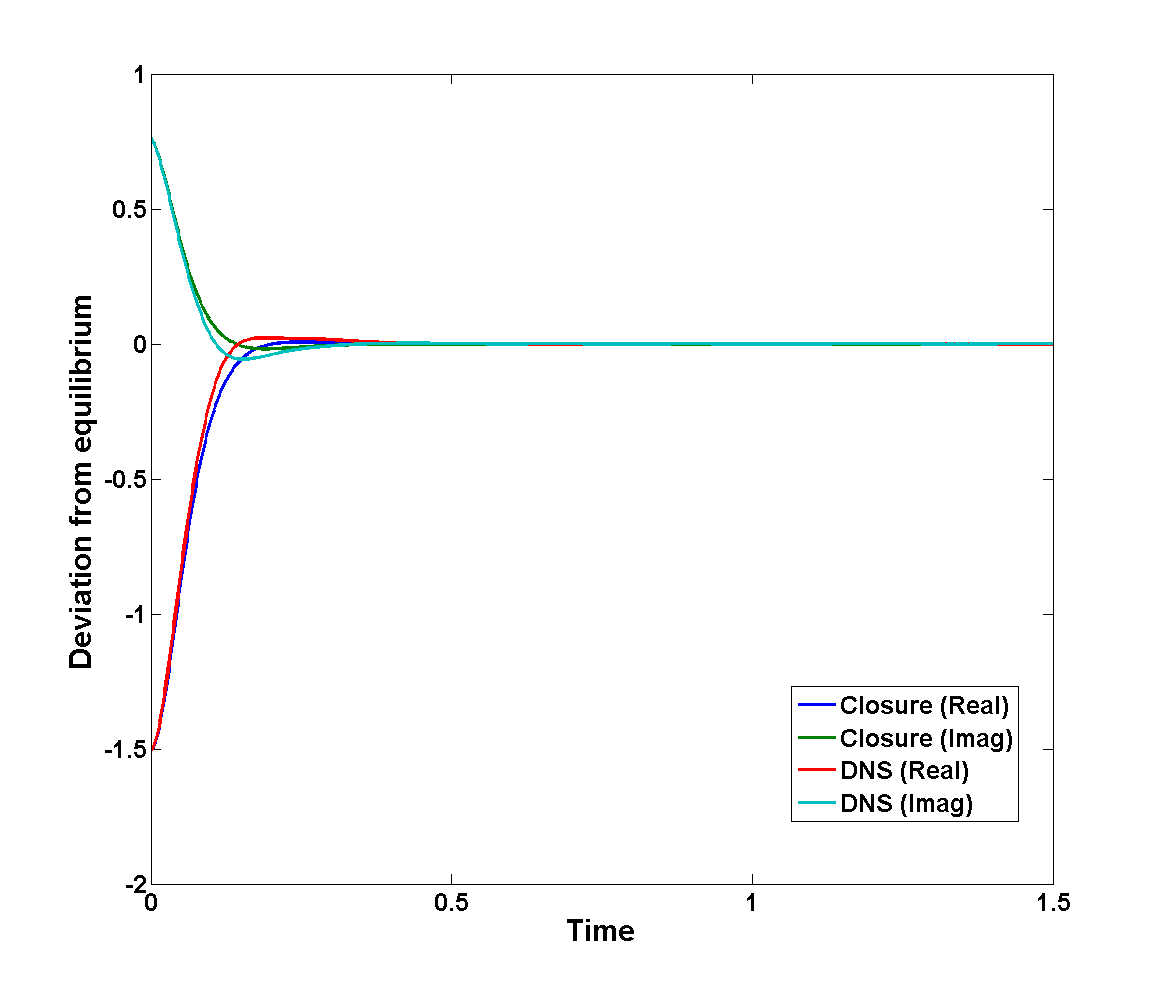}                
                \caption{Complex Mode 5}
                \label{fig:2e}
        \end{subfigure}
\caption{(continued)}
\label{Figure2}
\end{figure}
\subsection{Significantly removed from equilibrium case}

Historically, closures theories have  tended to be more successful
in situations that are near statistical equilibrium. 
 In the previous
subsection our comparisons used initial mean deviations smaller than a typical equilibrium excursion of
the system, namely,  $1/\sqrt{2\beta}$ for each real mode.  
We now,  therefore, pose a more stringent test of the closure to examine
the case significantly removed equilibrium behavior.   For simple expediency we multiply
all deviations in the initial mean resolved variables by a factor of $\sqrt{10}$, the $r_{dev}$ parameter. 
The initial condition amplitudes are therefore an order of magnitude larger than the first experiment.
This amplification  
also emphasizes the contribution of the quadratic advection terms in the closure
equations. The results
are displayed in Figure \ref{Figure2} in the same format as Figure \ref{Figure1}. The best
choice tunable parameter in this case is found to be $\gamma=73.83$, 
which corresponds to an exponential damping
time  of $0.2602/|k|$.  The
RMS error per mode and time step is  $0.02561$  in this instance.   
Since all curves have been scaled up by a factor of $10$,  this magnitude of error indicates
that the fit is approximately 25\% worse than in the close to equilibrium
case. 

The general behavior of the DNS equilibration is fairly similar
to the close to equilibrium case, but with some notable differences. In
particular, the general equilibration time scale is very much the
same as the close to equilibrium case, again reinforcing the success of
the reduced model with fractional diffusion. The primary difference is in
the plateau period, during which larger differences
between the closure and the DNS are evident,  particularly for the
lowest two wavenumbers (Figures \ref{fig:2a} and \ref{fig:2b}). This deviation suggests that the modification
of  the advection terms in the closure model is somewhat
less successful than the robust prediction of effective, fractional
diffusion.

\subsection{Other initial condition deviations from equilibrium}
We also examined the case $r_{dev}=1$ which lies midway between the choices considered above. Results are qualitiatively very similar
(and indeed slightly improved) over the close to equilibrium case. The best choice for damping time here was $0.28156/|k|$ which is
little removed from the close to equlibrium case considered above.  

In contrast the case $r_{dev}=10$ produces poor results. This, of course, corresponds with a very large, ten standard deviation, excursion from equilibrium. The RMS error per mode and time step is $0.235$ and this is around
four times what a simple rescaling of the results above would suggest. In addition the best fit damping time was dramatically shortened to $0.13805/|k|$,
about half what the previous three settings give. Qualitatively  the results (not shown) had big discrepancies
which were most apparent as the various modes approached equilibrium, when large oscillations in the DNS were not reproduced by the closure.
Failure of the closure in this case is perhaps not a complete surprise given that a Taylor series expansion about equilibrium was used to solve the Hamilton-Jacobi equation (see Section 6).   

\subsection{Robustness of tuning parameter}

The results above show that for the three settings, $r_{dev}=1/\sqrt{10}$, $r_{dev}=1$ and  
$r_{dev}=\sqrt{10}$, the damping time only varies by approximately $8\%$.
Given that the amplitude had been increased by an order of magnitude,
the small change in the single adjustable parameter $\gamma$ is a rather
reassuring indication of the robustness of the closure theory.   To illustrate this robustness further
we set the adjustable parameter $\gamma$ for the significantly removed from equilibrium case to be
equal to the best-fit parameter for the close to equilibrium case.   The ensuing closure dynamics is
exhibited in Figure \ref{Figure3}, which should be compared with Figure \ref{Figure2}.    

\begin{figure}
        \begin{subfigure}[b]{0.73\textwidth}
                \centering
                \includegraphics[width=\textwidth]{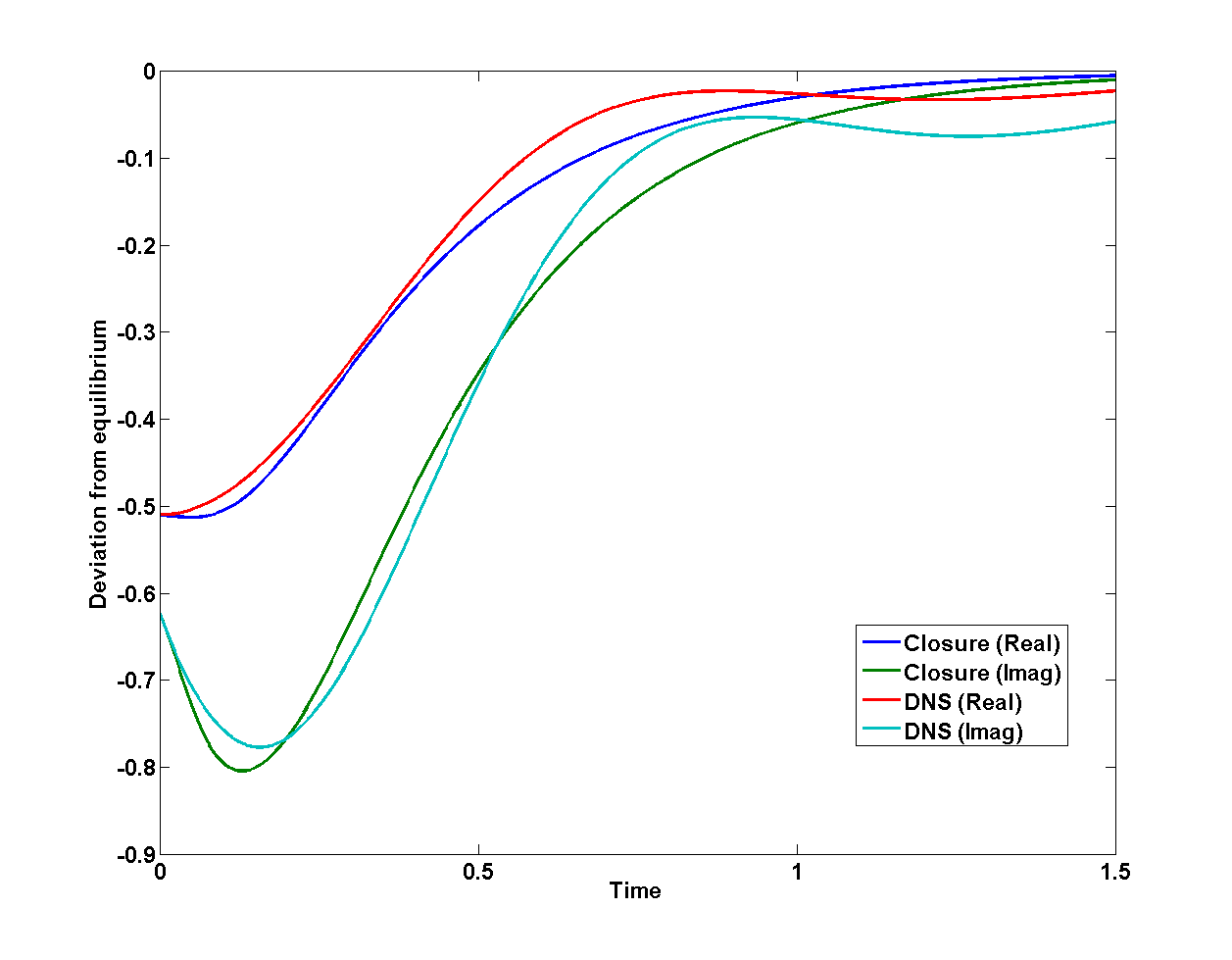}                
                \caption{Complex Mode 1}
                \label{fig:3a}
        \end{subfigure}

        \begin{subfigure}[b]{0.73\textwidth}
                \centering
                \includegraphics[width=\textwidth]{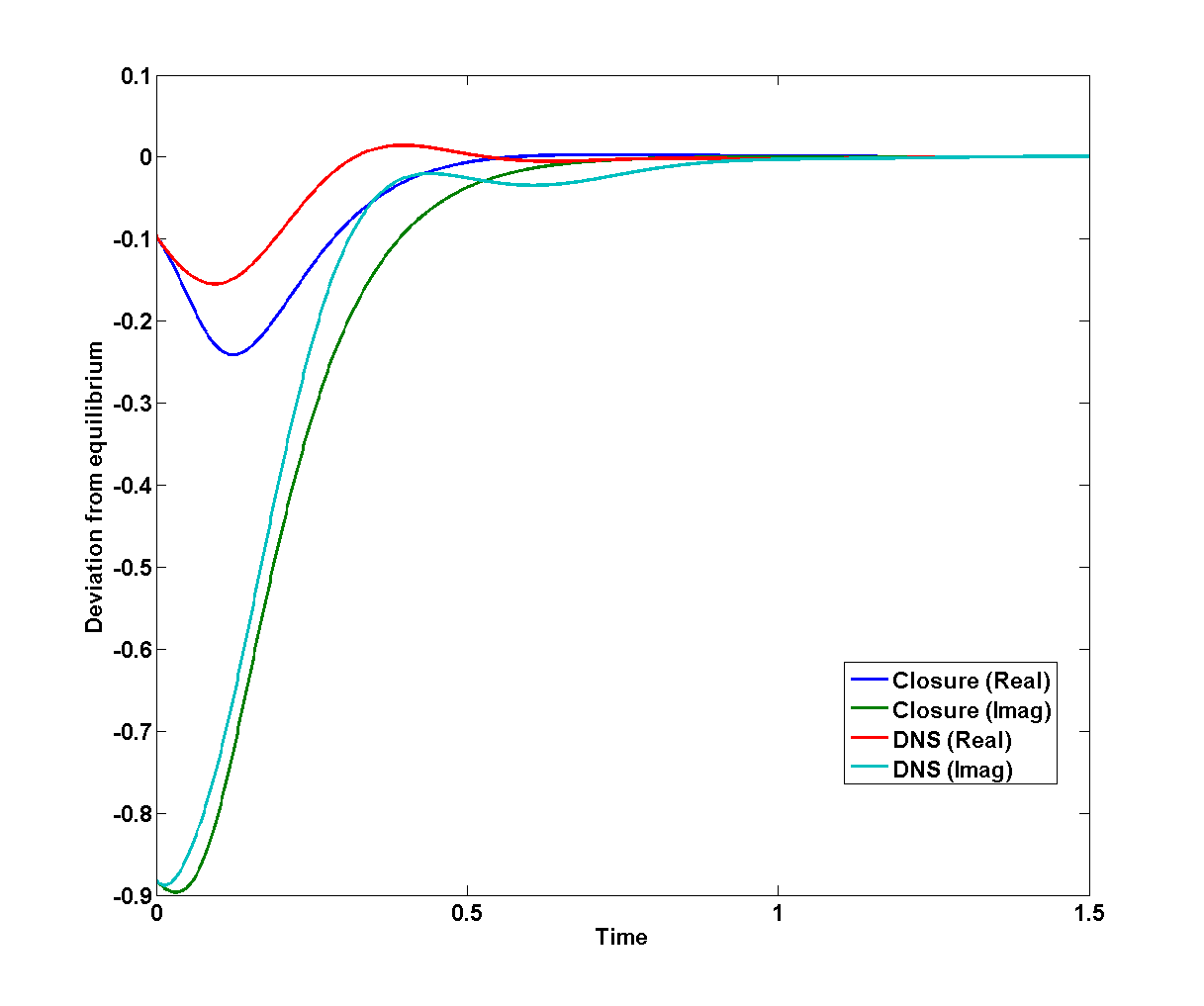}                
                \caption{Complex Mode 2}
                \label{fig:3b}
        \end{subfigure}
   \caption{Same as Figure \ref{Figure2} except that the tuning parameter is set to that applying for Figure \ref{Figure1} (see text). This tests the robustness of the tuning
parameter. This Figure should be compared with Figure \ref{Figure2}.}
\label{Figure3}
\end{figure}

\begin{figure}
     \ContinuedFloat 
        \begin{subfigure}[b]{0.73\textwidth}
                \centering
                \includegraphics[width=\textwidth]{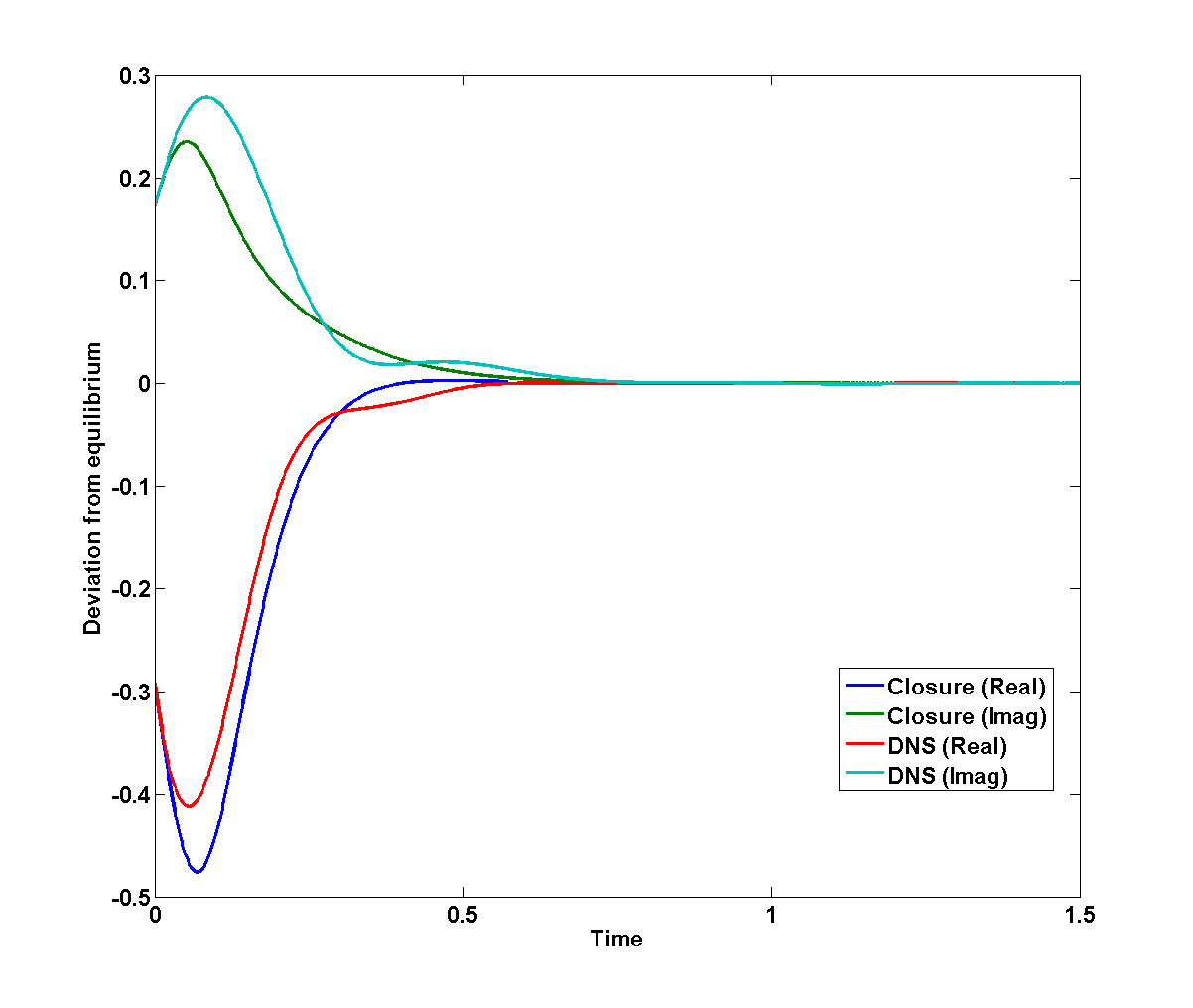}                
                \caption{Complex Mode 3}
                \label{fig:3c}
        \end{subfigure}

        \begin{subfigure}[b]{0.73\textwidth}
                \centering
                \includegraphics[width=\textwidth]{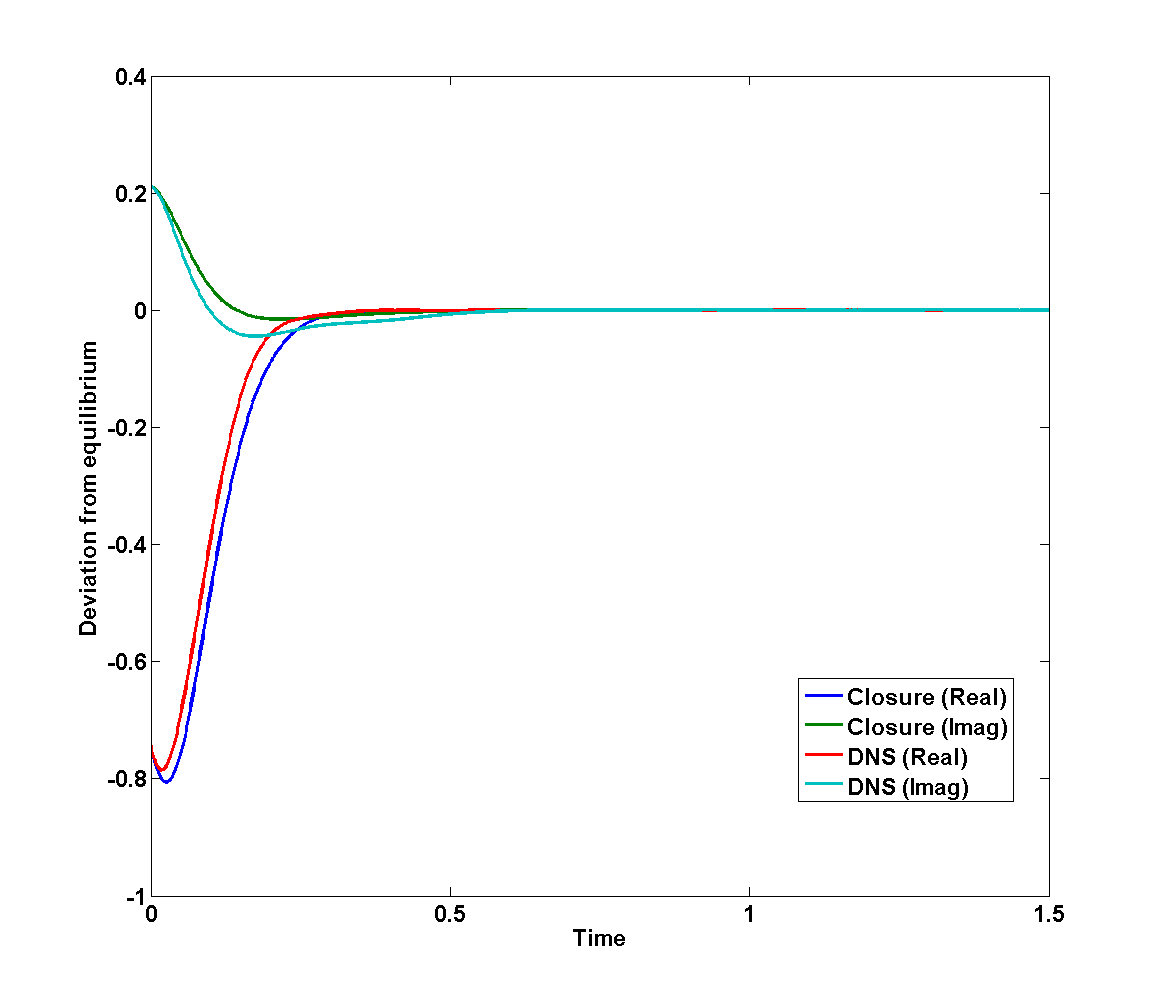}                
                \caption{Complex Mode 4}
                \label{fig:3d}
        \end{subfigure}
\caption{(continued)}
\label{Figure3}
\end{figure}
\begin{figure}
     \ContinuedFloat 

        \begin{subfigure}[b]{0.73\textwidth}
                \centering
                \includegraphics[width=\textwidth]{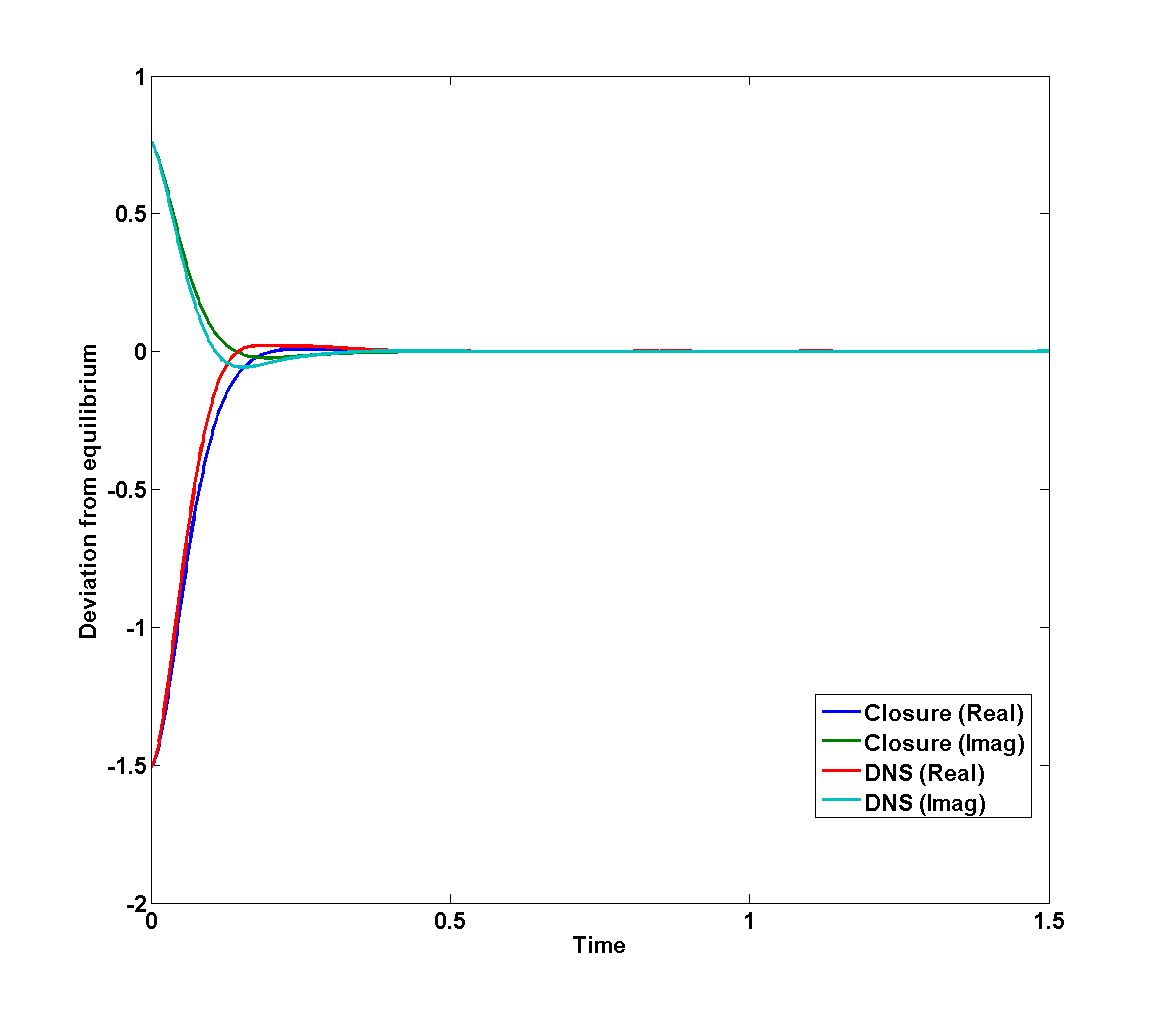}                
                \caption{Complex Mode 5}
                \label{fig:3e}
        \end{subfigure}
\caption{(continued)}
\label{Figure3}
\end{figure}

\subsection{A time scale separation experiment}

As noted previously the system under study does not exhibit a clear
separation of time scales between the slow, low wavenumber modes and
the ``neglected'' fast, high wavenumber modes. Indeed, as shown in
\cite{mati01}, this time scale is empirically observed to
be inversely proportional to wave number, just like the dissipation
time scale discussed above. Thus, for example, the second set of $2m$
modes have a time scale on average of only one third less than that
of the last retained mode. This issue makes the construction of closure
models for this system a challenge, as has been noted in detail in the
stochastic mode reduction work of Majda, Vanden-Eijnden and collaborators
(see \cite{mtv3}). Given this background  it is reasonable to suspect
that the discrepancies observed
above, most notably for the significantly removed from equilibrium case, are due in part
to time-scale separation issues.   To test this idea in the context of
our approach, we extend the closure model by incorporating an
additional $2m$ ``buffer'' modes in the resolved variables, to separate the 
time scale of original closure modes from that of the neglected ``heat bath''
modes. In this new configuration the first unresolved mode has
a time scale of one half of the last original resolved mode. 

The buffered results for significantly removed from equilibrium initial conditions  may
be seen in Figure \ref{Figure4} in the same format as the previous two experiments. Interestingly,
the best fit closure now exhibits a longer damping time of $0.3004/|k|$.
This slower equilibration is perhaps explained by the inclusion of the buffer modes which
mediate the nonlinear dissipation via the advection terms of the closure.
The fit of the original modes is significantly improved relative to the unbuffered run,
with the RMS error reduced to $0.0218$. Qualitatively, this improvement occurs most
strikingly in the higher wavenumber modes $4$ and $5$ ,where the
fit is now really quite impressive. This agreement might be expected
since presence of the buffer modes removes the abrupt discontinuity in the original model at
wavenumber $5$. There are also
some smaller improvements with the low wave-number modes, especially in the
plateau phase, but still
the discrepancy with respect to the DNS oscillations on approach to
equilibrium remains.

It is interesting to compare the results above to those in \cite{majda2006stochastic} where a stochastic reduction of the TBH system is proposed.
These authors consider the case in which $m=1,2$ and assess performance using auto-correlation rather than first moment as is done here.
In the near equilibrium case these are comparable due to the fluctuation dissipation relation. Like here they find
that relaxation to equilibrium is exponential with a decay time given by the eddy turnover time. They also find the DNS
oscillatory behaviour we observe and are also unable to produce it with their reduced model. They attribute this discrepancy to the lack of time separation between fast and slow modes. Our results with a significant buffer do not produce a large improvement in performance of the low wavenumber variables. This difference  deserves closer future attention.   
\begin{figure}
        \begin{subfigure}[b]{0.73\textwidth}
                \centering
                \includegraphics[width=\textwidth]{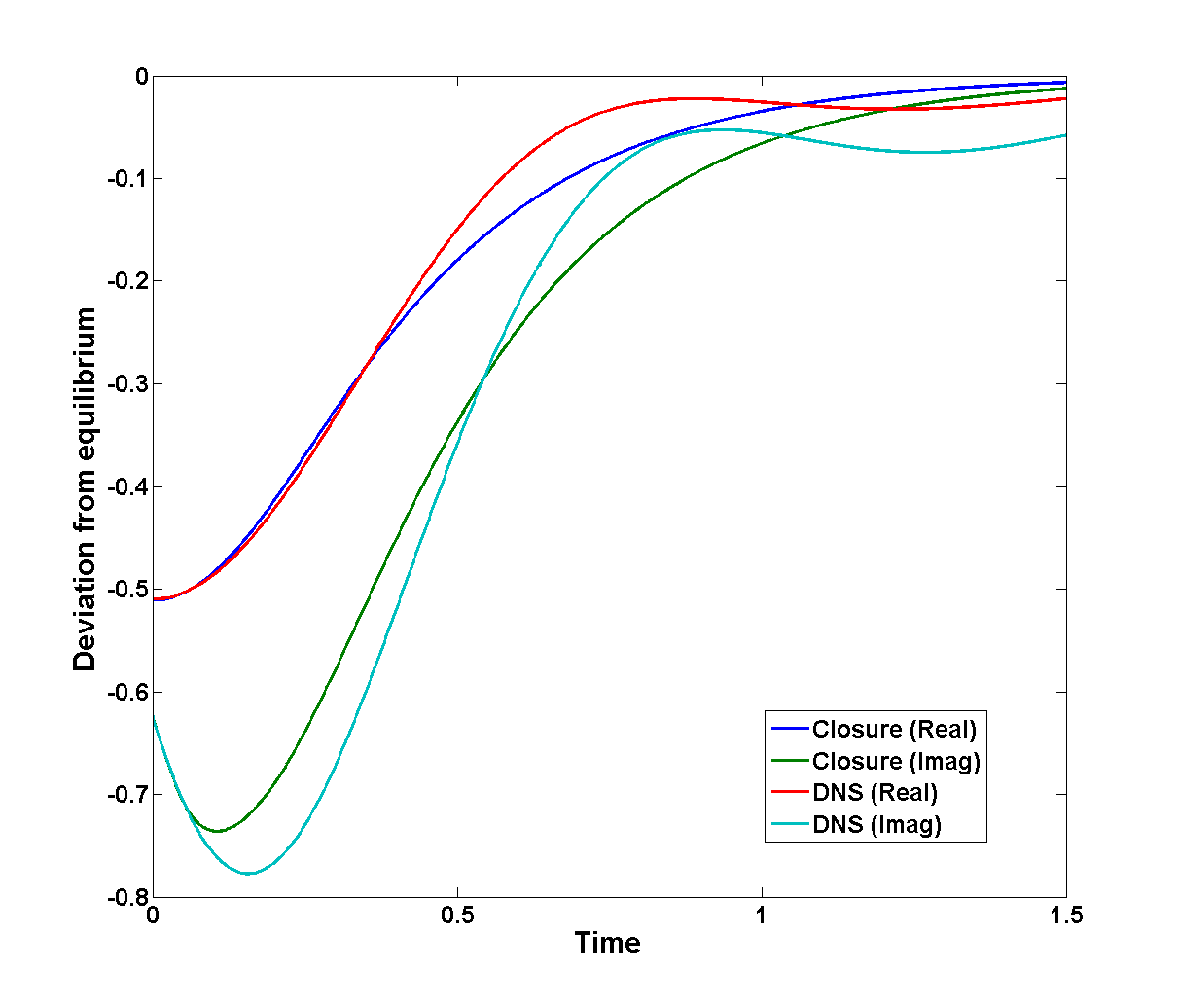}                
                \caption{Complex Mode 1}
                \label{fig:4a}
        \end{subfigure}

        \begin{subfigure}[b]{0.73\textwidth}
                \centering
                \includegraphics[width=\textwidth]{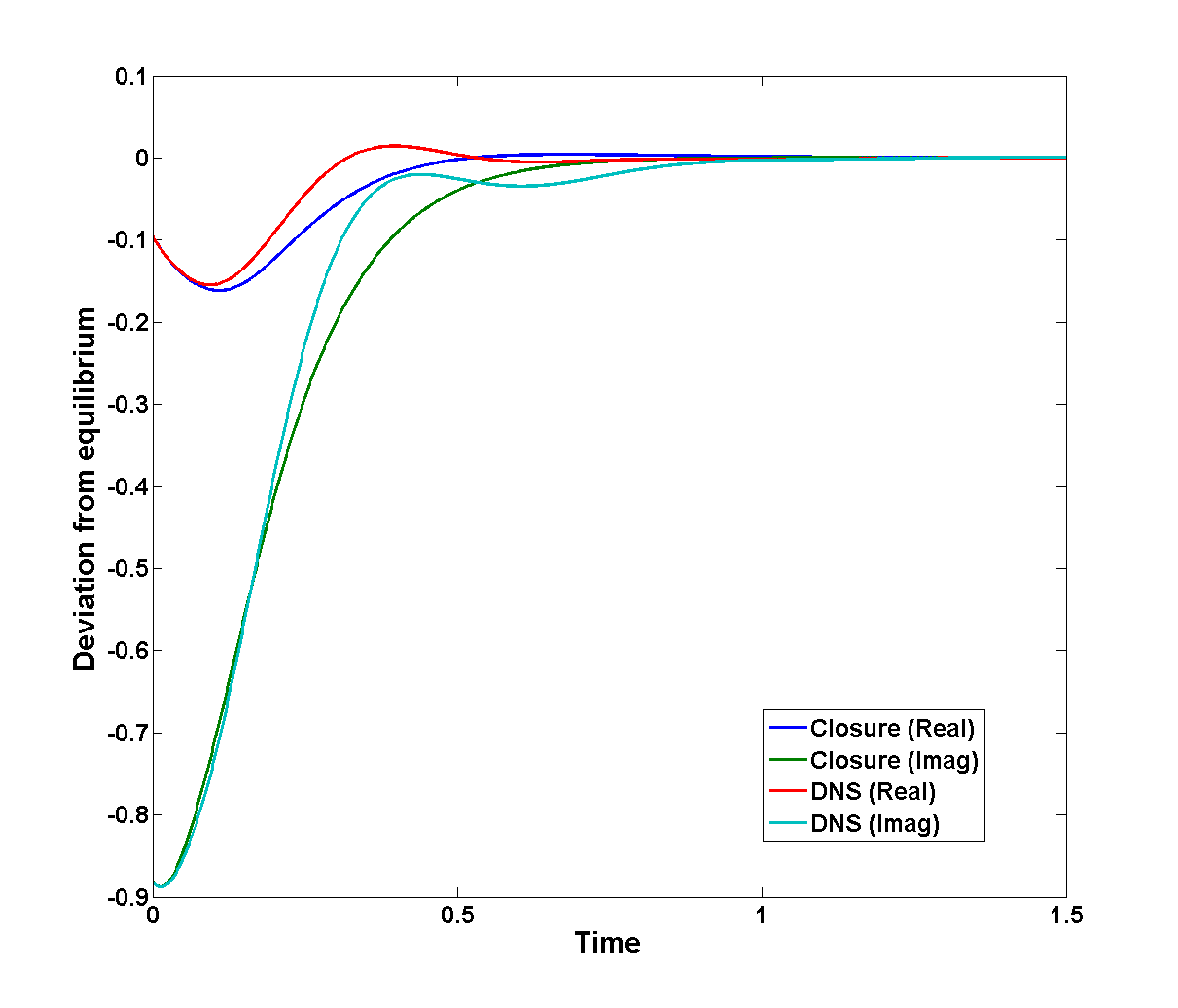}                
                \caption{Complex Mode 2}
                \label{fig:4b}
        \end{subfigure}
   \caption{Same as Figure \ref{Figure2} except the closure model has an additional 5 complex "buffer" modes (see text)}
\label{Figure4}
\end{figure}
\begin{figure}
     \ContinuedFloat 
        \begin{subfigure}[b]{0.73\textwidth}
                \centering
                \includegraphics[width=\textwidth]{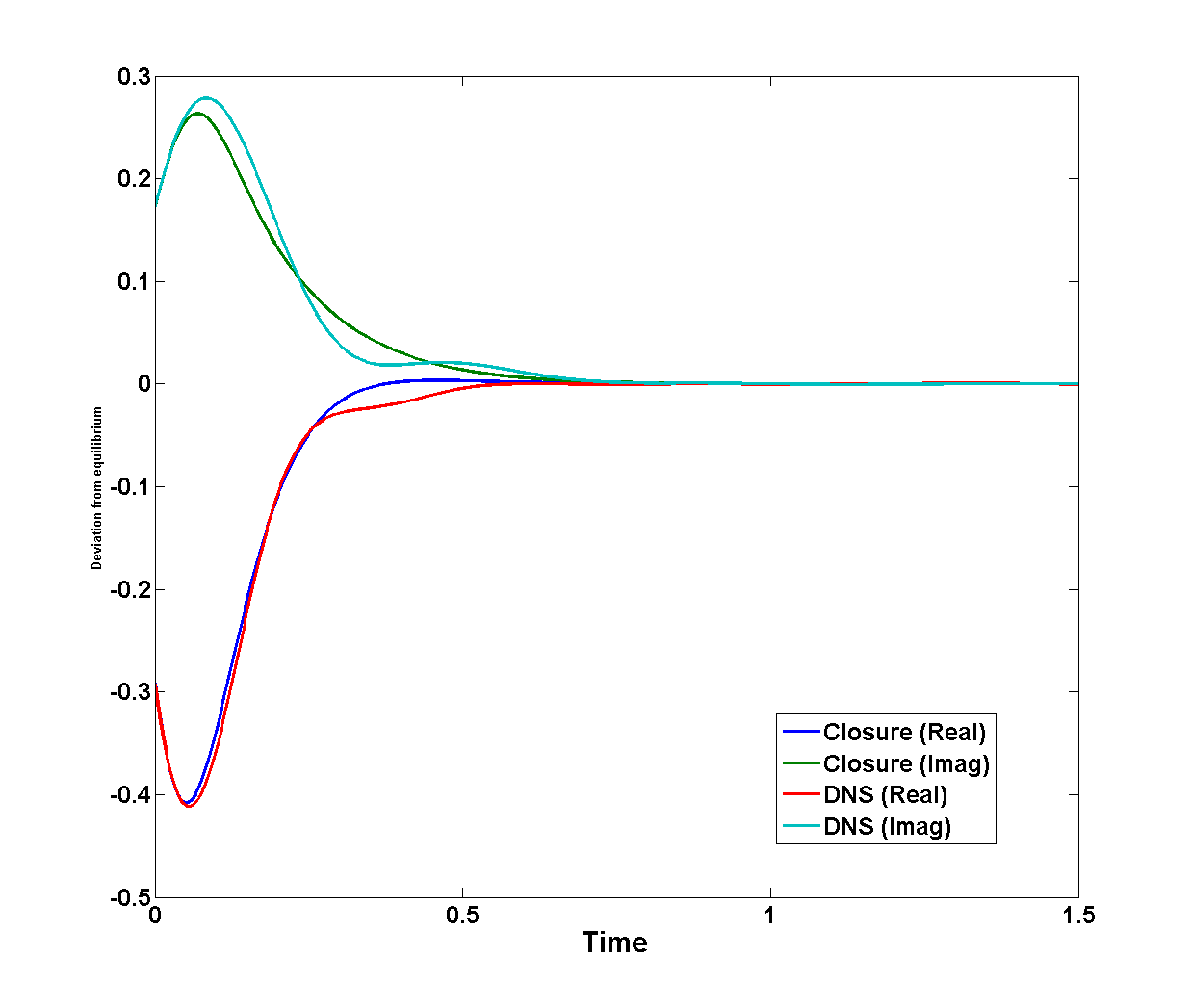}                
                \caption{Complex Mode 3}
                \label{fig:4c}
        \end{subfigure}

        \begin{subfigure}[b]{0.73\textwidth}
                \centering
                \includegraphics[width=\textwidth]{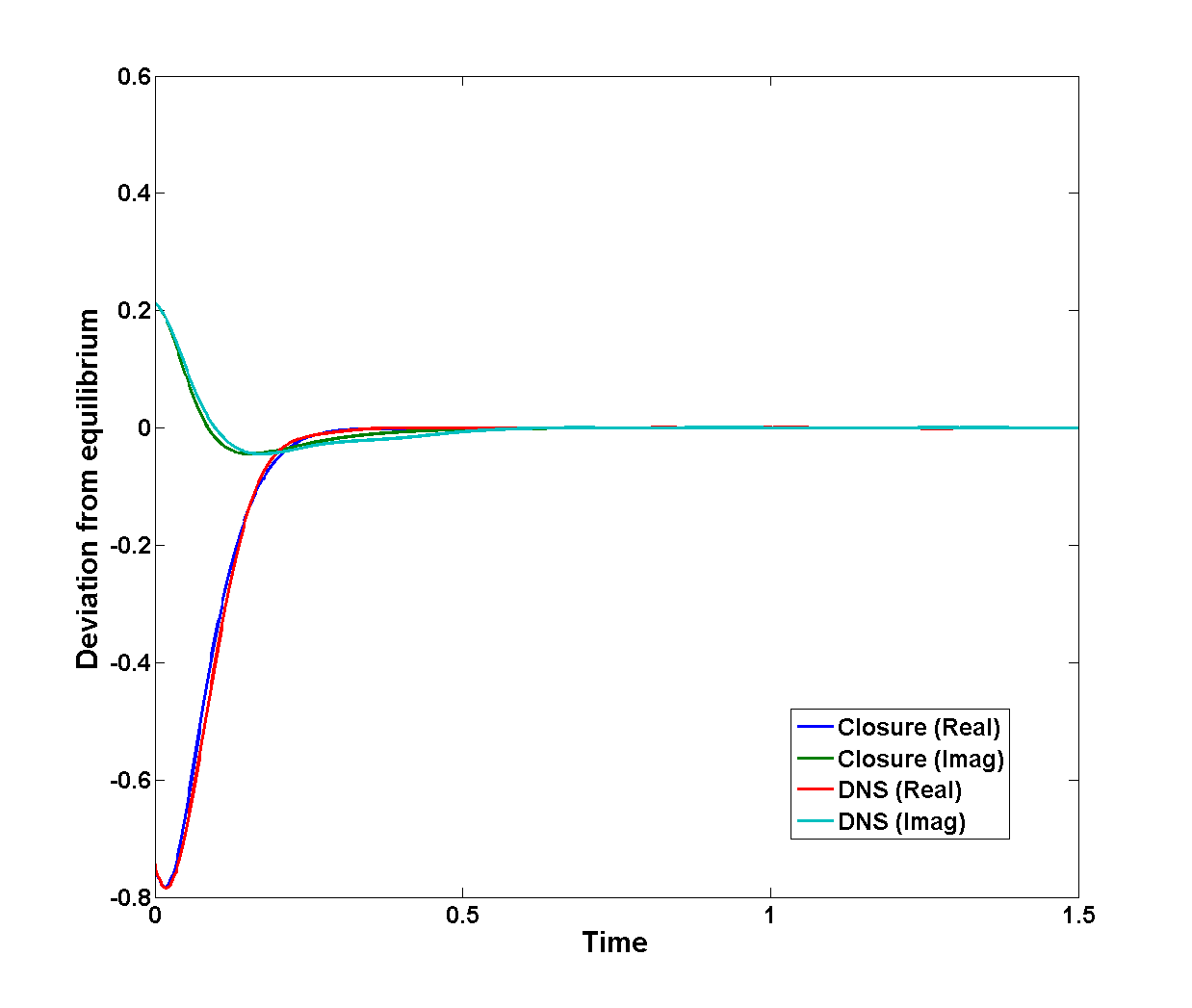}                
                \caption{Complex Mode 4}
                \label{fig:4d}
        \end{subfigure}
\caption{(continued)}
\label{Figure4}
\end{figure}
\begin{figure}
     \ContinuedFloat 
        \begin{subfigure}[b]{0.73\textwidth}
                \centering
                \includegraphics[width=\textwidth]{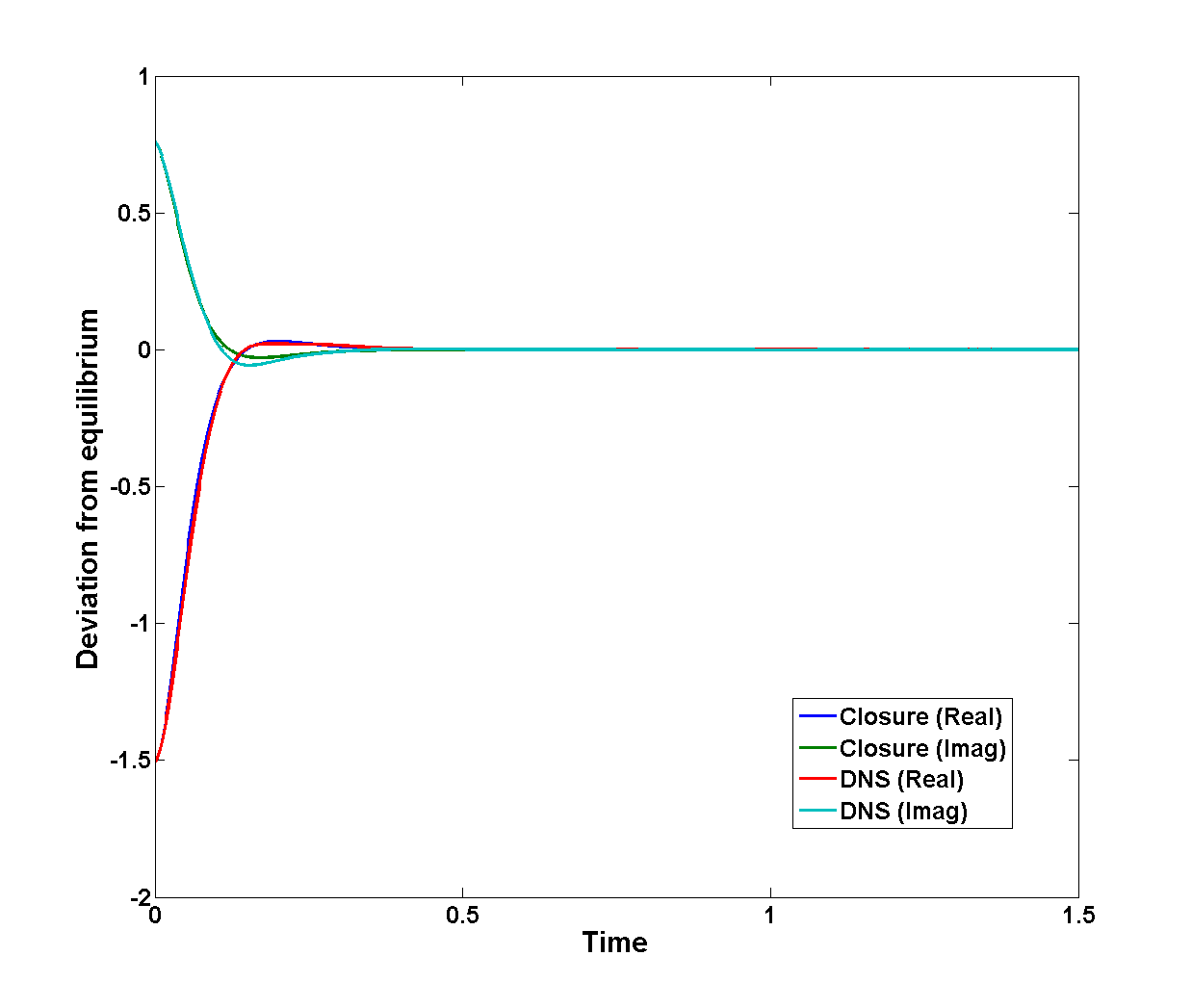}                
                \caption{Complex Mode 5}
                \label{fig:4e}
        \end{subfigure}
\caption{(continued)}
\label{Figure4}
\end{figure}

\section{Summary and Conclusions} 

A fundamental problem in nonequilibrium statistical mechanics, studies of turbulence
and stochastic modelling of complex systems is the  formulation of tractable reduced models of the 
slow, coherent part of the dynamical system. One of the authors has proposed a general variational principle for producing such closures in the context of Hamiltonian dynamics. 
This closure theory fits a continuous time series of trial probability densities to the Liouville equation
using a weighted, time-integrated, mean-squared cost functional to quantify the lack-of-fit.  
The closed reduced equations satisfied by minimizers of the cost functional are then obtained
by classical Hamilton-Jacobi theory.   More details may be found in the companion paper \cite{Tur12}.

In this contribution we have applied this closure theory to a Galerkin-truncated Burgers-Hopf model,
 which has been proposed as a simple analog of more complex and realistic fluid systems. 
The model is governed by a Hamiltonian dynamics, has a Gibbs invariant 
measure,  and exhibits mode decorrelation with time scales that vary inversely with wavenumber.
These properties make it an ideal testbed for future work with more realistic turbulent systems. 
In the present work we have concentrated on the free relaxation of the system from nonequilibrium statistical
initial conditions.    

Our results pertain to a reduced model for the lowest $5$ complex modes in a dynamical
system of $50$ nonlinearly interacting complex modes.    In light of the simplicity of the Gibbs measure,
under which all the modes are independent Gaussians, it is possible to carry out the calculation
of the governing equations for the reduced model explicitly up to second-order (quadratic nonlinearity)
in the mean resolved variables.    The result is a set of governing equations for the temporal evolution of the first moments of the resolved modes which resembles a severe truncation of  original 
nonlinear system with the following modifications and characteristics.

\begin{enumerate}

\item There is a fractional dissipation which is proportional to the wavenumber of the mode. 
Remarkably this fractional diffusion 
is the minimum such regularization required to prevent finite-time singularities in the continuous
(infinite mode) Burgers-Hopf system \cite{KNS}. 
It is the primary mechanism by which the reduced model irreversibly equilibrates, and 
as expected by linear-response theory  it has the same wavenumber-dependent time scale as the modal decorrelation time.

\item The nonlinear terms of the best-fit reduced model have modified coefficients which imply an altered energy flux between resolved modes compared with the original TBH system.   By means of these coefficients  the reduced model approximates the indirect influence of the unresolved modes on the interactions 
between the mean resolved modes.    

\item Both these modifications of the dynamics are time-dependent in that they take a time comparable to the equilibration time to manifest themselves in full. This initial period, which we term the plateau time, is incorporated naturally in the nonstationary version of our closed reduced model.     
 
\end{enumerate}

The first-moment best-fit closure model has been compared with 
a direct numerical simulation (DNS) of the full TBH system using very large
ensembles. 
The comparisons are generally good for a broad range of initial deviations from equilibrium. 
If however the deviation is
made sufficiently large the fit does break down significantly. 
This is not surprising given that the closure
equations rely on a Taylor expansion of the Hamilton-Jacobi equation centered on equilibrium. 
 
The wavenumber-dependent equilibration time scale predicted
by the best-fit closure matches 
that produced by the DNS across the spectrum of resolved modes.     
Only one scalar parameter is adjusted empirically to fit the closure predictions to the DNS results. 
The value of this parameter is observed to be quite robust even as the imposed nonequilibrium
initial conditions are increased by an order of magnitude.   

The TBH dynamics is a rather severe test for reduced models given that there is no clear separation of timescales. 
We have found that the performance of the reduced model can be improved by inserting some ``buffer'' modes, 
which widens the separation of time scales between the slow and fast parts of the system.     Nonetheless, some slow oscillatory behavior of the low resolved modes is missed by
the reduced model, indicating perhaps that there are memory effects inherent in the statistical dynamics.
It is conceivable that an extended set of resolved variables that includes the time derivatives of
the lowest modes might be able to capture those oscillations.

Given that the TBH dynamics has been developed as 
a paradigm for more complex dynamical systems having quadratic nonlinearities typical
of hydrodynamical equations of motion, the question arises whether the methodology that we
have applied in the present investigation could be extended to other systems.  
When coarse-graining complex systems of this type onto 
their low, slow modes, one may expect to encounter  Liouville 
residuals having similar structure to that in the TBH case.    
Under such circumstances cost functions akin to the one that 
we have constructed in the TBH case, which quantify 
the collective effect of products of pairs of modes,  
may prove to be an efficacious way to represent the influence of unresolved
fluctuations on the reduced dynamics.   
Future investigations on other systems of this kind exhibiting
turbulent dynamics  would be helpful in evaluating the range of applicability of our 
best-fit approach and  the useful forms of the cost functions that define it.      
The reader is referred to \cite{Tur12} for some further discussion on these issues.

        

\appendix
     
\section{Ancilliary Material}   

Here we collect some standard conventions and calculations 
concerning functions of several complex variables which 
may be useful to the reader a various points in the paper.   

For any smooth, complex-valued function $f(z)$, of $n$ complex variables, 
$z =(z_1 , \ldots , z_n) \in \C^n$, with $z_k=x_k + i y_k$, the usual derivatives are defined by 
\[
\frac{\d f}{ \d z_k}  = \half  \left( \,  \frac{\d f }{ \d x_k}  - i \frac{\d f}{ \d y_k}  \, \right) \, , \;\;\;\;\;\;\;\; 
\frac{\d f}{ \d z_k^*}  = \half  \left( \,  \frac{\d f}{ \d x_k}  + i \frac{\d f}{ \d y_k}  \, \right) \, ;
\]
the notation $z_k^* = x_k -i y_k$ is used for complex conjugate.    In terms of these,
the chain rule for the composite function $f(z(t))$, where $t$ is a real variable, is simply
\[
\frac{d }{ d t}  f(z(t)) =  \sum_{k=n}^n 
    \frac{\d f}{ \d z_k} \frac{d z_k}{d t} +  \frac{\d f}{ \d z_k^*} \frac{d z_k^*}{ d t}  \, 
              =   \,   \sum_{k=-n}^n   \frac{\d f}{ \d z_k} \frac{d z_k}{d t}  \, .
\]
The last equality uses the  convention that $z_{-k} = z_k^*$, which 
occurs naturally in the context of a Fourier representation of a dynamics in
terms of complex amplitudes $z_k$.

Repeated application
of this chain rule to the function $f(tz)$, $0 \le t \le1$, yields the complex form of Taylor's
expansion,
\[
f(z) = f(0) +  \sum_{k=-n}^n L_k \, z_k  +   \sum_{k_1,k_2=-n}^n  M_{k_1 k_2} \,  z_{k_1} z_{k_2}
                  +   \sum_{k_1,k_2, k_3=-n}^n  N_{k_1 k_2 k_3} \, z_{k_1} \, z_{k_2}z_{k_3}
                       +  O( |z|^4)  \, ,
\]
with coefficients 
\[
 L_k =   \frac{\d f}{ \d z_k}(0) \, , \;\;\;\; 
 M_{k_1 k_2} =  \half  \frac{\d^2 f}{ \d z_{k_1} \d z_{k_2}}(0)  \, , \;\;\;\;
 N_{k_1 k_2 k_3}  = \frac{1}{6}   \frac{\d^3 f}{ \d z_{k_1} \d z_{k_2} \d z_{k_3}} (0)  \, .  
\]
In Section 5  this expansion is used to determined an approximation to the value function,
$v(\lambda)$, for $\lambda \in \C^m$, and in Section 6 for the time-dependent
value function $v(\lambda,t)$.    For those calculations this form of the expansion is 
better than the equivalent multi-index form.  

The Lagrangian $\L(\lambda,\lambdadot)$ and the Hamiltonian $\H(\lambda,\mu)$ that
occur in the defining optimization principle and the associated Hamilton-Jacobi equation,
respectively, are related by the Legendre tranform (\ref{legendre}).    It is instructive to
relate this complexified Legendre transform to the familiar real transform.    
Write the complex variables as $\lambda_k = \xi_k + i \eta_k$ and  
$\mu_k = \phi_k + i \psi_k$,  for $k = 1, \ldots, m$.    Then,  (\ref{legendre}) is equivalent to
\[
\phi_k = \frac{\d }{ \d \dot{\xi}_k }  \, \frac{\L}{2}  \, ,   \;\;\;\;   
  \psi_k = \frac{\d }{ \d \dot{\eta}_k }  \, \frac{\L}{2}  \, ,  \;\;\;\;
  \frac{\H}{2}  =    \sum_{k=1}^m \phi_k \dot{\xi}_k  + \psi_k \dot{\eta}_k \, - \, \frac{\L}{2} \, . 
\]
Thus the complex Legendre transform (\ref{legendre}) between $\L$ and $\H$ 
is identical with the real tranform between $\L/2$ and $\H/2$.
The stationary value function, $v(\lambda)$, in (\ref{value-fn}) is defined by 
the action integral for $\L(\lambda,\lambdadot)$ and satisfies the complex
form of the Hamilton-Jacobi equation (\ref{HJ}).  
The conjugacy relation (\ref{conjugate-relation})  determined by $v$ is 
\[
\mu_k  =    - \frac{\d v}{\d \lambda_k^*}(\lambda)  = 
      - \left[ \,  \frac{\d }{ \d \xi_k } + i  \frac{\d }{ \d \eta_k }  \, \right] \frac{v}{2}   \, ,
\]
This relation shows that (\ref{HJ}) and (\ref{conjugate-relation}) are identical with
the real Hamilton-Jacobi theory
applied to the Hamiltonian $\H(\xi, \eta, \phi, \psi) /2$ and the value function $v(\xi, \eta)/2$.  
The nonstationary case is complexified in the same way.       The common factor 
of $1/2$ throughout is irrelevant to the analysis, being absorbed in the complex variable
conventions.





\ack 

In the course of this work  the authors benefited from conversations
with  A.~J. Majda and E. Vanden Eijnden.     
Part of this research was completed during B.T.'s 
two-month visit to the Courant Institute of Mathematical Sciences
in Spring 2012. The authors also acknowledge the use of the Burgers-Hopf numerical model code 
authored by I. Timofeyev.


\frenchspacing
\bibliographystyle{cpam}
\bibliography{bruce,refs}

\begin{thebibliography}{10}
\providecommand{\url}[1]{\texttt{#1}}
\providecommand{\urlprefix}{Available at: }
\providecommand{\eprint}[2][]{\url{#2}}

\bibitem{abramov2003hamiltonian}
Abramov, R.; Kovacic, G.; Majda, A. Hamiltonian structure and statistically
  relevant conserved quantities for the truncated {B}urgers-{H}opf equation.
  \emph{Communications on pure and applied mathematics} \textbf{56} (2003),
  no.~1, 1--46.

\bibitem{arnol1989mathematical}
Arnol'd, V.~I. \emph{Mathematical methods of classical mechanics},
  Springer-Verlag, New York, 1989.

\bibitem{balescu1975equilibrium}
Balescu, R. \emph{Equilibrium and nonequilibrium statistical mechanics}, Wiley,
  New York, 1975.

\bibitem{bryson1975applied}
Bryson, A.~E.; Ho, Y.-C. \emph{Applied optimal control: Optimization,
  estimation, and control}, Hemisphere Pub. Corp.(Washington and New York),
  1975.

\bibitem{casella2001statistical}
Casella, G.; Berger, R.~L. \emph{Statistical inference}, Duxbury Press, 2001.

\bibitem{Cha87}
Chandler, D. \emph{Introduction to modern statistical mechanics}, Oxford
  University Press, 1987.

\bibitem{chorin2000optimal}
Chorin, A.~J.; Hald, O.~H.; Kupferman, R. Optimal prediction and the
  {M}ori--{Z}wanzig representation of irreversible processes. \emph{Proc. Nat.
  Acad. Sci.} \textbf{97} (2000), no.~7, 2968--2973.

\bibitem{chorin2002optimal}
Chorin, A.~J.; Hald, O.~H.; Kupferman, R. Optimal prediction with memory.
  \emph{Physica D: Nonlinear Phenomena} \textbf{166} (2002), no.~3, 239--257.

\bibitem{chorin1998optimal}
Chorin, A.~J.; Kast, A.~P.; Kupferman, R. Optimal prediction of underresolved
  dynamics. \emph{Proc Nat. Acad. Sci.} \textbf{95} (1998), no.~8, 4094--4098.

\bibitem{de2011non}
De~Groot, S.~R.; Mazur, P. \emph{Non-equilibrium thermodynamics}, North
  Holland, 1962.

\bibitem{evans1998partial}
Evans, L.~C. Partial differential equations. graduate studies in mathematics
  19. \emph{American Mathematical Society}  (1998).

\bibitem{fleming1975deterministic}
Fleming, W.~H.; Rishel, R.~W. \emph{Deterministic and stochastic optimal
  control}, vol. 268, Springer-Verlag New York, 1975.

\bibitem{gelfand2000calculus}
Gelfand, I.~M.; Fomin, S.~V. \emph{Calculus of variations}, Dover publications,
  2000.

\bibitem{givon2004extracting}
Givon, D.; Kupferman, R.; Stuart, A. Extracting macroscopic dynamics: model
  problems and algorithms. \emph{Nonlinearity} \textbf{17} (2004), no.~6, R55.

\bibitem{katz1967principles}
Katz, A. \emph{Principles of statistical mechanics: the information theory
  approach}, WH Freeman, 1967.

\bibitem{keizer1987statistical}
Keizer, J. \emph{Statistical thermodynamics of nonequilibrium processes},
  Springer, 1987.

\bibitem{KNS}
Kisalev, A.; Nazaraov, F.; Shterenberg, R. Blow up and regularity for fractal
  {B}urgers equation. \emph{pre-print}  (2008). ArXiv:0804.3549.

\bibitem{klmati02}
Kleeman, R.; Majda, A.~J.; Timofeyev, I. Quantifying predictability in a model
  with statistical features of the atmosphere. \emph{Proc. Nat. Acad. Sci. USA}
  \textbf{99} (2002), 15\,291--15\,296.

\bibitem{kullback1997information}
Kullback, S. \emph{Information theory and statistics}, Dover publications,
  1997.

\bibitem{lanczos1986variational}
Lanczos, C. \emph{The variational principles of mechanics}, Dover Publications,
  1986.

\bibitem{lax2005hyperbolic}
Lax, P.~D. Hyperbolic systems of conservation laws ii. \emph{Selected Papers
  Volume I}  (2005), 233--262.

\bibitem{luzzi2002predictive}
Luzzi, R.; Vasconcellos, {\'A}.~R.; Ramos, J.~G. \emph{Predictive Statistical
  Mechanics: A Nonequilibrium Ensemble Formalism}, Springer, 2002.

\bibitem{mtv3}
Majda, A.; Timofeyev, I.; Vanden-Eijnden, E. A priori tests of a stochastic
  mode reduction strategy. \emph{Physica D} \textbf{170} (2002), 206--252.

\bibitem{majda2006stochastic}
Majda, A.; Timofeyev, I.; Vanden-Eijnden, E. Stochastic models for selected
  slow variables in large deterministic systems. \emph{Nonlinearity}
  \textbf{19} (2006), no.~4, 769.

\bibitem{majda2010mathematical}
Majda, A.~J.; Harlim, J.; Gershgorin, B. Mathematical strategies for filtering
  turbulent dynamical systems. \emph{Discrete and Continuous Dynamical Systems}
  \textbf{27} (2010), no.~2, 441--486.

\bibitem{mati00}
Majda, A.~J.; Timofeyev, I. Remarkable statistical behavior for truncated
  {B}urgers-{H}opf dynamics. \emph{Proc. Nat. Acad. Sci. USA} \textbf{97}
  (2000), 12\,413--12\,417.

\bibitem{mati01}
Majda, A.~J.; Timofeyev, I. Statistical mechanics for truncations of the
  {B}urgers-{H}opf equation: a model for intrinsic stochastic behavior with
  scaling. \emph{Milan Journal of Mathematics} \textbf{70(1)} (2002), 39--96.

\bibitem{mtv2}
Majda, A.~J.; Timofeyev, I.; Vanden-Eijnden, E. A mathematics framework for
  stochastic climate models. \emph{Comm. Pure Appl. Math.} \textbf{54} (2001),
  891--974.

\bibitem{ottinger2005beyond}
{\"O}ttinger, H.~C. \emph{Beyond equilibrium thermodynamics},
  Wiley-Interscience, 2005.

\bibitem{sagan1992introduction}
Sagan, H. \emph{Introduction to the Calculus of Variations}, Dover
  Publications, 1992.

\bibitem{Tur12}
Turkington, B. An optimization principle for deriving nonequilibrium
  statistical models of {H}amiltonian dynamics. \emph{pre-print}  (2012).
  ArXiv:1207.2692.

\bibitem{Zub74}
Zubarev, D.~N. \emph{Nonequilibrium {S}tatistical {T}hermodynamics}, Plenum
  Press, New York, 1974.

\bibitem{zwanzig2001nonequilibrium}
Zwanzig, R. \emph{Nonequilibrium statistical mechanics}, Oxford University
  Press, USA, 2001.

\end{thebibliography}

\end{document}